\documentclass[useAMS,usenatbib]{mn2e}
\usepackage{graphics}
\usepackage{supertabular}

\usepackage{tabularx}

\usepackage{multirow}

\usepackage{epic}

\usepackage{rotating}

\usepackage{subfigure}

\usepackage{aas_macros}

\usepackage{xcolor}

\usepackage{amsmath}
\newcommand{\OHI}{$\Omega_{\rm HI}$}

\newcommand{\ObHI}{$\Omega_{\rm HI}b_{\rm HI}$}
\newcommand{\tcm}{21cm}
\newcommand{\fht}{f_{\rm H2}}
\newcommand{\Msun}{$ M_{\odot}$}
\newcommand{\Ms}{ M_{\odot}}
\newcommand{\hM}{ h {\rm Mpc^{-1}}}

\newcommand{\Mpch}{ h^{-1}\rm Mpc}

\title[Intensity Mapping Cross-Correlations]{Intensity Mapping Cross-Correlations: Connecting the Largest Scales to Galaxy Evolution}
\author[L.Wolz et al.]{L. Wolz$^{1, 2}$\thanks{E-mail:
lwolz@unimelb.edu.au}, C. Tonini$^{1}$, C. Blake$^{3}$ , J.S.B. Wyithe$^{1, 2}$ \\
$^{1}$School of Physics, University of Melbourne, Parkville, VIC 3010, Australia\\
$^{2}$ARC Centre of Excellence for All-Sky Astrophysics (CAASTRO)\\
$^3$Centre for Astrophysics \& Supercomputing, Swinburne University of Technology, P.O. Box 218, Hawthorn, VIC 3122, Australia\\
}
\begin{document}

\date{}

\pagerange{\pageref{firstpage}--\pageref{lastpage}} \pubyear{}

\maketitle

\label{firstpage}

\begin{abstract}
Intensity mapping of the neutral hydrogen (HI) is a new observational tool that can be used to efficiently map the large-scale structure of the Universe over wide redshift ranges. The power spectrum of the intensity maps contains cosmological information on the matter distribution and probes galaxy evolution by tracing the HI content of galaxies at different redshifts and the scale-dependence of HI clustering. The cross-correlation of intensity maps with galaxy surveys is a robust measure of the power spectrum which diminishes systematics caused by instrumental effects and foreground removal. 
We examine the cross-correlation signature at redshift $z\approx 0.9$ using a variant of the semi-analytical galaxy formation model SAGE \citep{Croton:2016} applied to the Millennium simulation in order to model the HI gas of galaxies as well as their optical magnitudes based on their star-formation history. We determine the scale-dependent clustering of the cross-correlation power for different types of galaxies determined by their colours, which act as a proxy for their star-formation activity. We find that the cross-correlation coefficient with HI density for red quiescent galaxies  falls off more quickly on smaller scales $k>0.2\hM$ than for blue star-forming galaxies. Additionally, we create a mock catalogue of highly star-forming galaxies using a selection function to mimic the WiggleZ survey, and use this to predict existing and future cross-correlation measurements of the Green Bank telescope and Parkes telescope. We find that the cross-power of highly star-forming galaxies shows a higher clustering on small scales than any other galaxy type and that this significantly alters the power spectrum shape on scales $k>0.2\hM$. We show that the cross-correlation coefficient is not negligible when interpreting the cosmological cross-power spectrum. On the other hand, the cross-correlation coefficient contains information about the HI content of the optically selected galaxies.
\end{abstract}
\begin{keywords}
cosmology -- galaxy evolution --radio astronomy.
\end{keywords}

\section{Introduction}
Intensity Mapping of neutral hydrogen (HI) is a new observational tool to map the large scale structure of the Universe at various redshifts. Galaxies emit radio waves at \tcm~due to hyperfine structure of atomic hydrogen present in cold gas reservoirs. The \tcm~spectral line has been detected in galaxies in the local Universe \citep{2003AJ....125.2842Z,2010ApJ...723.1359M} and also used to map the structure of galaxies \citep{Walter:2008bq}. However, due to the weakness of the signal the detection is challenging at higher redshifts. Intensity mapping aims to observe the unresolved line emission of HI integrated over each frequency channel, tracing the large-scale hydrogen distribution on angles greater than the telescope beam;  typically of order of tens of arc minutes at target redshifts $z\approx 1$. Furthermore, it probes the entire HI mass function and, thus, does not suffer any selection effects. Intensity Mapping has been proposed as a very efficient tool to map the clustering of galaxies \citep{2004MNRAS.355.1339B, 2009astro2010S.234P} and to measure the Baryon Acoustic Scale as a probe of the Cosmic expansion \citep{Chang:2007xk, Wyithe:2007rq} for a wide range of redshifts. There are two ways of conducting observations; via single dish telescopes such as the Green Bank telescope (GBT) \citep{Chang:2010jp} and via interferometric arrays such as CHIME \citep{Bandura2014}  or the future Square Kilometre Array (SKA) \citep{Santos:2015vi}. The GBT team has attempted to statistically detect the intensity mapping signal at $z\approx 0.8$ \citep{Chang:2010jp, 2013MNRAS.434L..46S}. \cite{2013ApJ...763L..20M} reported a detection of the \tcm~signal in the GBT data via the cross-correlation with optical galaxies. There are on-going observations at the GBT with median redshift 0.8 to increase the signal-to-noise of the previous datasets. In addition, the Parkes telescope is targeting the same fields with a median redshift of $z \approx 0.9$.

The challenges in observing intensity maps are manyfold. On the one hand, any instrumental fluctuations of the telescope can cause significant systematics in the continuum maps. This implies that observations require robust calibration, accurate beam shapes and pointings, and negligible polarization leakage. Inhomogeneous thermal noise of the telescope, which can be similar in amplitude to the expected \tcm~signal, can pose additional complexities. On the other hand, terrestrial and astrophysical foregrounds can significantly contaminate the observations. Radio interference by terrestrial, human signals can introduce spatial and frequency dependent contaminations in the maps.  The foregrounds from our own Galaxy predominantly originate from synchrotron emission which, at the wavelengths of interest, can be of 4-5 magnitudes higher than HI signal.  Additionally, extra-galactic point sources contaminate the maps. The issues of foreground removal have been discussed using a variety of parametric and blind methods to subtract foregrounds, and applied to simulations, i.e. \cite{2014arXiv1401.2095S,2014MNRAS.441.3271W}. It is widely acknowledged that the removal of the foregrounds poses the major challenge in present and future intensity mapping experiments. 

Cross-correlation of the intensity maps with galaxy surveys has been suggested in order to beat the systematic errors caused by instruments and foregrounds, and to increase the statistical significance of the detection (e.g. \citealt{Villaescusa-Navarro:2014rra,2015arXiv150903286P}). The analysis of the GBT data presented in \cite{2013ApJ...763L..20M} cross-correlated the HI maps with galaxies observed by the WiggleZ Dark Energy survey \citep{Drinkwater:2009ev} which are selected as star-forming galaxies. It is crucial for the cosmological analysis to model an accurate prediction of the amplitude and the shape of the cross-correlation power spectrum.

The amplitude of the HI power spectrum depends on the HI abundance in our Universe which is poorly constrained by current observations for redshifts higher than 0.1. An additional factor is the HI bias, determined by the distribution of the HI relative to the underlying dark matter field. When considering the cross-correlation with galaxy surveys, the power is further dictated by the bias of the optically selected galaxies and the cross-correlation coefficient. This coefficient is determined by the intrinsic correlation between the HI and selected galaxies, and is sensitive to the amount of HI gas present in the optical galaxies. 

In this work, we simulate the cross-correlation of HI intensity maps and optical galaxies by applying a semi-analytical model (SAM) to N-body simulations for a box at $z\approx0.9$. We use two variations of star-formation recipes to model the galaxy evolution, which also predict the amount of neutral hydrogen by splitting the cold gas into atomic and molecular phases. We compute the photometric emission of galaxies using their star-formation history, and divide them into quiescent and star-forming galaxy populations depending on their observed colour. We verify that the relation between star-formation activity and HI abundance of the galaxies is observable in the cross-correlation. We show that the cross-correlation coefficient exhibits a strong scale-dependence for different galaxy selections and thus significantly influences the shape of the cross-correlation power spectrum. Conversely, this result implies that the shape of the cross-correlation coefficient contains information about the star-formation history and HI content of the optically selected galaxies. In addition, we specifically model a galaxy selection mimicking the WiggleZ survey, to provide theoretical predictions for present and future observations.

The article is structured as follows. In Sec.~\ref{sec2}, we present the simulation details including an outline of star-formation recipes, modelling of HI intensity maps and galaxy photometries. Furthermore, we describe the selection criteria for the mock WiggleZ selection. Sec.~\ref{sec3} contains the results of the galaxy selections applied to the SAM. We present and discuss the auto and cross power spectra of the simulation in Sec.~\ref{sec4} for the different galaxy colours. We conclude in Sec.~\ref{sec5}.

\section{Simulation Details}
\label{sec2}
\begin{figure*}
\subfigure[All galaxies]{
\includegraphics[scale=0.33]{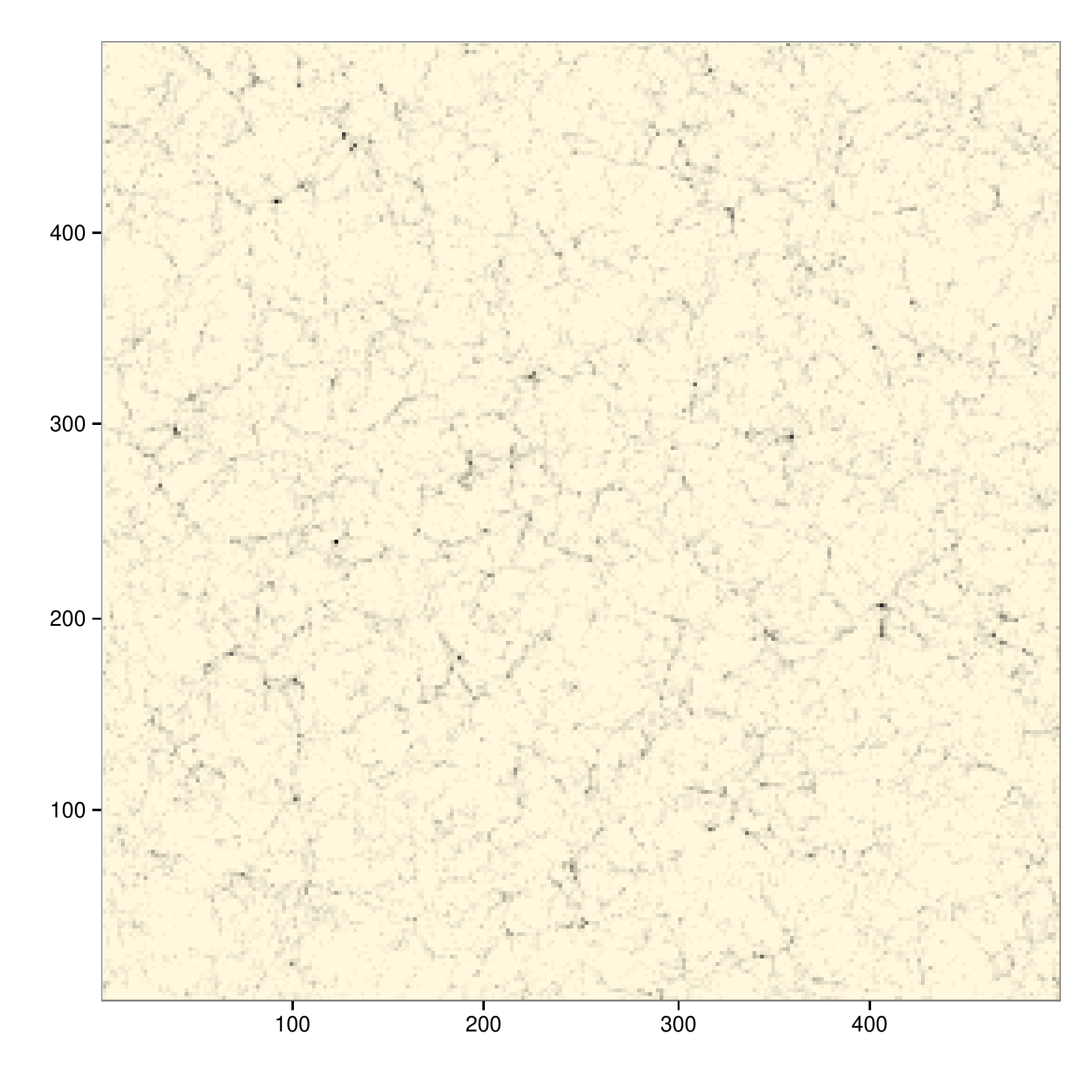}
\label{SAGEall}
}\subfigure[HI intensity map]{
\includegraphics[scale=0.33]{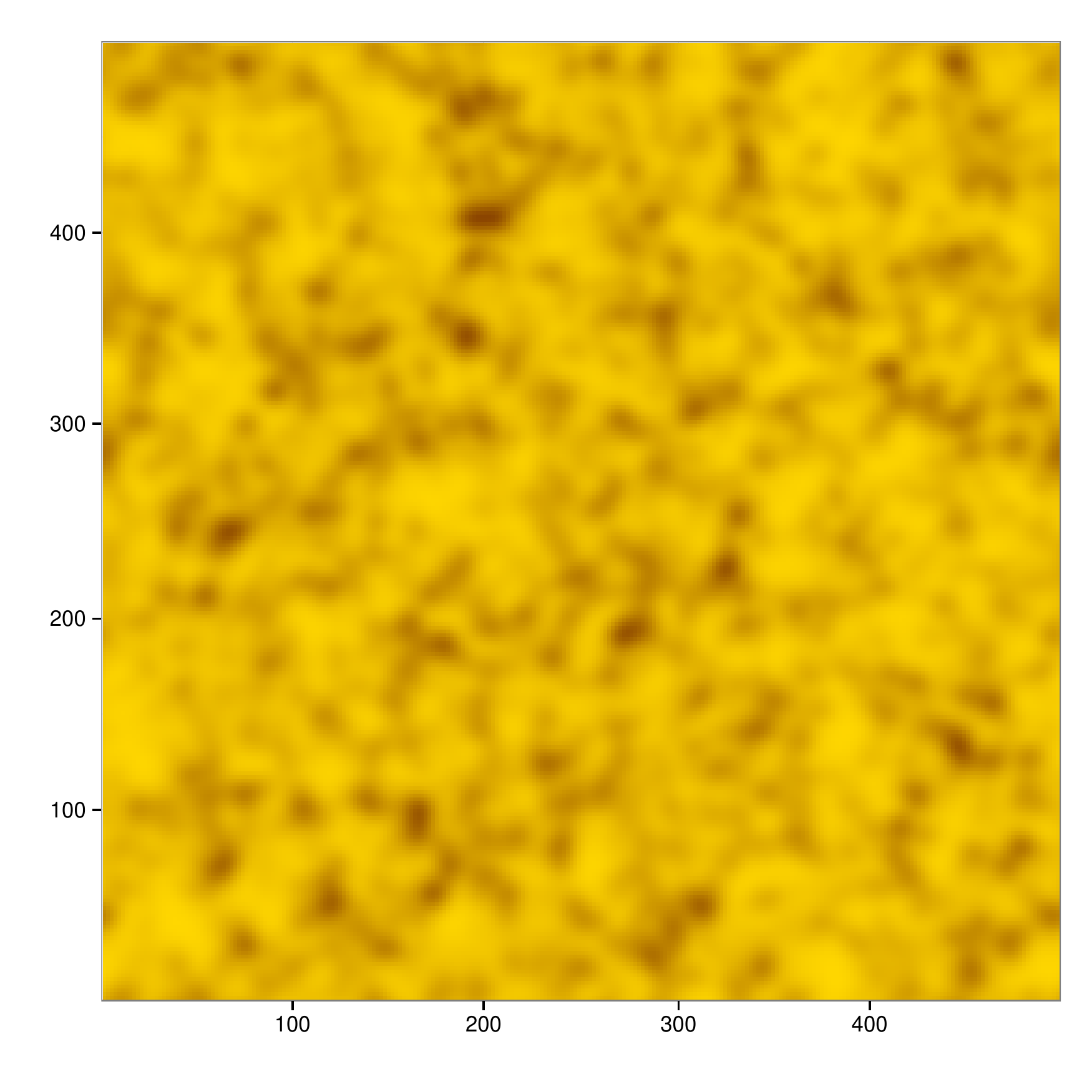}
\label{SAGEHI}
}\subfigure[WiggleZ galaxies]{
\includegraphics[scale=0.33]{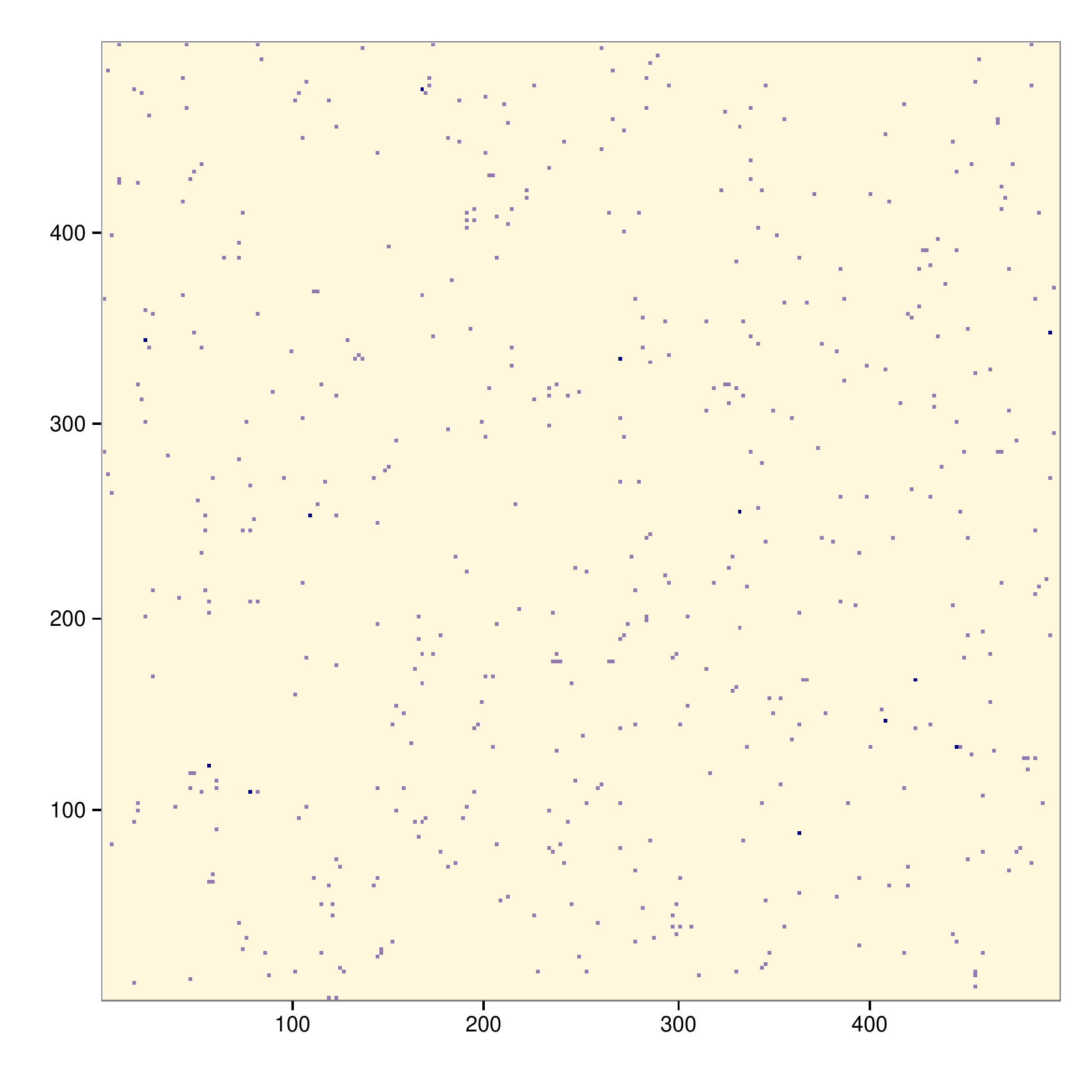}
\label{SAGEWiggleZ}
}
\caption{Maps of a slice through the simulation box with line-of-sight width of approximately $2\Mpch$ for the KD simulation.}
\label{maps}
\end{figure*}
In our work, we make use of a variation of the semi-analytical galaxy formation model SAGE (based on \citealt{2006MNRAS.365...11C,Croton:2016}; see also \citealt{Tonini:2016}) run on the Millennium simulation \citep{2005Natur.435..629S} with a comoving volume of $(500\Mpch)^3$ and particle mass resolution of $8.6\times10^8h^{-1}$ \Msun. The semi-analytic model assigns the universal baryon fraction to each halo in the form of hydrogen, shock-heated to the halo virial temperature. The model follows the baryonic physics inside each halo, which includes: gas infall into the halo, cooling of the hot gas into a cold disky component, star formation and supernova feedback, AGN feedback, metal enrichment, galaxy mergers, disk instabilities, the formation of spheroidal stellar components, gas stripping and star formation quenching, and gas outflows (for a detailed description of each physical recipe, see \citealt{2006MNRAS.365...11C,Croton:2016}).

In the following, we briefly describe the star formation recipes which determine the HI gas and star formation histories of each galaxy.  An example of the galaxy distribution of our simulation is displayed in Fig.~\ref{SAGEall}.

\subsection{Star Formation Models}
\label{SFmodels}
\cite{Croton:2016} implemented the commonly used star formation law established by \cite{RobertCKennicutt:1997id} which relates the star-formation rate surface density to the gas surface density as $\Sigma_{\rm SFR}=A \Sigma_{\rm gas}^N$ through a power law where the normalisation factor $A$ and $N$ are empirically determined. This relation has been shown to break for the outer regions of spirals and for dwarf galaxies. Furthermore, this law uses a critical surface density $\Sigma_{\rm crit}(r)$ below which star formation is suppressed.  

In our simulation, we use two star-formation models which differ from the standard SAGE star formation recipe in order to split the cold gas into molecular and atomic phases.

\begin{itemize}
\item{Krumholz-Dekel (KD model)\\ 
The star-formation recipe in \cite{2012ApJ...753...16K} (hereafter KD) relates the star formation rate $\dot M_*$ to the molecular gas density scaled by the free fall time of the gas $t_{\rm ff}$. The fraction of molecular hydrogen $\fht$ in the galaxy is determined by the surface density and the metallicity $Z$ of the galaxy. These two properties are assumed to approximate the main mechanisms controlling the phases of the cold gas. Molecular hydrogen formation is increased by collisionally-excited metal line cooling and a higher abundance of dust grains which act as a catalyst to the formation of molecular hydrogen. The molecules can be destroyed into their atoms through UV heating processes.

The relation of the molecular gas fraction to the metallicity can be approximated as 
\begin{equation}
\fht\sim \Sigma/(\Sigma + 10Z_0^{-1}M_{\odot} \rm pc^{-2})
\end{equation}
with $\Sigma$ the total gas surface column density and $Z_0$ the normalised metallicity. This implies, for regions where the column density is larger than the metallicity factor given in the equation above, that the gas is mainly in molecular phase and vice versa. 

The star formation rate (SFR) is  fuelled by the molecular gas of the galaxy and can be approximated by a linear relation 
\begin{equation}
\dot M_*=2\pi\int_0^{\infty} \fht\frac{\epsilon_{\rm ff}}{t_{\rm ff}}\Sigma_{g} r \rm{d}r
\label{SFRKD}
\end{equation}
where $\Sigma_{\rm g}$ is the gas surface density and the ratio of the star-formation efficiency to free-fall time is approximated as $\epsilon_{\rm ff}/{t_{\rm ff}} \propto (2.6{\text{ Gyr}})^{-1} (\Sigma_{\rm g}/0.18)^\gamma$ with $\gamma=\pm 0.33$ depending on $\Sigma_{\rm g}$ smaller or greater than 0.18. The integral is performed over the radius $r$ of the halo and all variables within the integral are dependent on the scale of the halo. For the details of the implemented expression, we refer the reader to Equs.(10-21) in  \cite{2012ApJ...753...16K}.
}
\item{Blitz-Rosolowsky (BR model)\\
The second recipe employed in our work is based on the description by \cite{Blitz:2006fh} (hereafter BR) which relates the molecular fraction of the hydrogen to the gas pressure of the disk. The star formation rate depends similarly as in  Equ.~\ref{SFRKD}  on the gas density, the molecular hydrogen fraction and the star-formation timescale which is, in this case, determined by the molecular gas depletion time as in \cite{2011ApJ...730L..13B}. In this model, the molecular gas fraction $\fht$ is determined by the hydrostatic pressure $P_{\rm ext}$ as
\begin{equation}
\fht=1/\left(1+\left(\frac{P_{\rm ext}}{P_0}\right)^{-\alpha}\right)
\end{equation}
where $P_0$ and $\alpha$ are parameters which depend on the stellar surface density, the gas density and the vertical velocity dispersion of the gas. This star formation description deviates from the Kennicutt-Schmidt law for molecule-poor galaxies. The details of this model can be found in \cite{Blitz:2006fh} Equs.(16-21).}

\item{AGN feedback}\\
We also test how "radio mode" AGN feedback changes the star-formation evolution and how this influences our analysis. We add AGN feedback into the KD star-formation model and label this case as KD AGN. The AGN feedback is implemented following the Bondi-Hoyle accretion model, as described in \cite{2006MNRAS.365...11C}. The central black hole accretes hot gas from the surrounding region delimited by the Bondi radius, which depends on the black hole mass and the speed of sound in the gas. The rate of accretion onto the black hole is proportional to the black hole mass and the dark matter halo virial temperature. This implementation suppresses gas cooling onto the most massive galaxies and contributes to the establishment of the red sequence and the drop of the galaxy stellar mass function at the high-mass end. \end{itemize}

\subsection{Intensity Maps}
We transform the SAGE output into intensity maps by assigning HI mass to each galaxy via $M_{\rm HI}=M_{\rm g}(1-\fht)$ where $M_{\rm g}$ is the cold gas mass. This can imply that halos without a resolved galaxy due to their low stellar mass are assigned a relatively high HI mass from the gas density. We assign the HI masses to a grid with number of pixels equal to $N^3=(256)^3$ and convert the gridded HI masses $M_{\rm HI}(x_i)$ into HI temperature per pixel $x_i$ via
\begin{equation}
T_{\rm HI}(x_i)=\frac{3A_{12}h_{\rm P}c^2}{32\pi m_{\rm H}k_{\rm B}\nu_{\rm 21}} \frac{M_{\rm HI}(x_i)}{\chi^2(\bar z) \Delta \nu(\bar z)  \Omega_{pix}},
\end{equation}
where  $h_{\rm P}$ is the Planck constant, $k_{\rm B}$ is the Boltzmann constant, $m_{\rm H}$ is the mass of
the hydrogen atom, $A_{12}$ the emission coefficient of the 21cm line transmission and $\nu_{\rm 21}$
the rest frequency of the 21cm emission.
 $\chi$ is the comoving distance to the redshift at medium redshift $\bar z$, which is given as snapshot redshift, and $\Omega_{\rm pix}$ is the surface area of the pixel in the perpendicular direction.
Intensity mapping experiments are naturally performed in redshift space and measure the surface brightness temperature of the HI emission integrated over a bandwidth $\Delta \nu$ at a given medium redshift $z$. For our simulation, this means that the cuboid data needs to be transformed into tomographic maps of HI temperature. To determine the bandwidth of the maps, we project the cube onto redshift space and determine the lower and upper redshift limits of the cube $(z^0, z^1)$ as well as the median redshift width  $\Delta z=(z_1-z_0)/N$. The median bandwidth in our settings is $ \Delta \nu =0.21\rm MHz$. We do not include the effects of peculiar velocities and focus on the real-space clustering in this study. 

We convolve the HI temperature maps with a telescope beam modelled as a symmetric, two-dimensional Gaussian function which only acts in the perpendicular directions of the data. The Gaussian beam is set up with a full width half maximum of $\theta_{\rm FWHM}=0.3 \rm deg$ which comparable to present single dish telescopes such as the Green Bank telescope or the Parkes telescope. The convolution is done using the convolution theorem by Fourier-transforming each intensity map $T_i$  and multiplying with the Fourier transformed beam before reverseing the Fourier transform to obtain maps in spatial direction. One example of an intensity map is shown in Fig.~\ref{SAGEHI}.

\subsection{Galaxy Photometry}
To calculate the galaxy luminosity, we employ the spectro-photometric model by \cite{2012ApJ...759...43T}; see also \cite{Tonini:2008hs,Tonini:2009nw}. The model records the star formation history along the merger tree of each galaxy from the semi-analytic model output, together with the metal content, and calculates the total galaxy spectrum using synthetic stellar populations; in the current work we use \cite{Conroy:2008mp,Conroy:2009gw}. The spectro-photometric model 
calculates apparent magnitudes in the GALEX UV and Sloan Digital Sky Survey g, r and i bands. 
The colour selection $\rm{NUV} - \rm{r}$ is tightly correlated with the star formation history of galaxies as shown in \cite{Salim:2004ct}. We select a star forming and a quiescent galaxy population based on their colours, using the selection criteria $( \rm{NUV} - \rm{r}) < 2$ for blue/star forming galaxies and $4 < (\rm{NUV} - \rm{r})$ for red/quiescent ones. 
\subsection{Case study: WiggleZ selection}
As a specific example, we model the photometric survey selection of the WiggleZ galaxy survey which has been used in the cross-correlation of intensity mapping data in \cite{2013ApJ...763L..20M}. The fields of the WiggleZ survey are also targeted in on-going intensity mapping experiments by the Parkes telescope.
 
The WiggleZ Survey  \citep{Drinkwater:2009ev} is a large-scale galaxy
redshift survey of bright emission-line galaxies over the redshift range
$z < 1$, with median redshift $z \approx 0.6$ and galaxy bias factor $b
\sim 1$.  The survey was carried out at the Anglo-Australian Telescope (AAT)
between August 2006 and January 2011.  In total $\sim 200{,}000$
redshifts were obtained, covering $\sim 1000$ deg$^2$ of equatorial sky
divided into seven well-separated regions. 

The photometric selection is 
\begin{equation}
\begin{split}
\rm{NUV}<22.8\\
20<\rm{r}<22.5\\
\rm{FUV}>23 \textrm{ or }(\rm{FUV}-\rm{NUV})>1\\
-0.5<(\rm{NUV-r})<2.
\end{split}
\label{Wigsel}
\end{equation}
The first two equations give the sensitivity limits of the GALEX Medium Imaging Survey and the Sloan Digital Sky Survey survey photometry. Additionally, the relations in the third line establishes the selection of Lyman-break galaxies for high redshifts. The fourth condition selects blue, star-forming galaxies. Since we are modelling high-redshift galaxies we do not need to apply the additional optical colour cuts used to exclude low-redshift galaxies from the WiggleZ sample.
The photometric selection is applied to the observed WiggleZ magnitudes which have been corrected for dust extinction as described in \cite{Drinkwater:2009ev}. One example of WiggleZ-selected galaxies is shown in Fig.~\ref{SAGEWiggleZ}.
%
%

\section{Study of Galaxy Properties}
We investigate the connection between star formation activity, HI abundance and galaxy colour as predicted by the SAGE simulation output for different star-formation recipes. We emphasise that for $z\approx0.9$ these galaxy properties are not detectable with current instrumentation, however, understanding the underlying relations will help with interpreting the observables, such as the power spectrum and can be used for future measurements. 
\label{sec3}

\subsection{HI Properties}
\begin{figure*}
\begin{tabular}{cc}             
\includegraphics[scale=0.5,  clip=true, trim= 0 0 0 50]{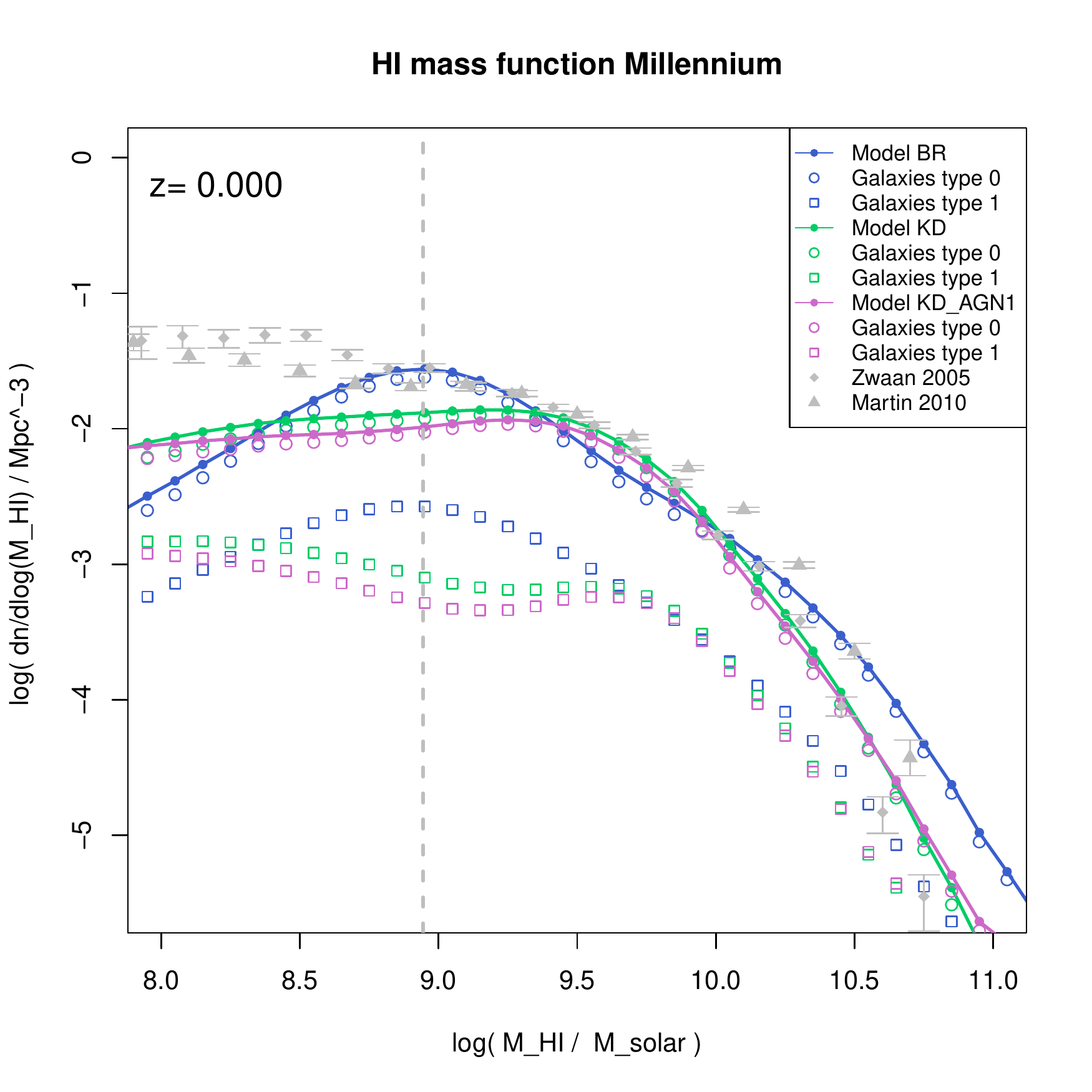}
\includegraphics[scale=0.5, clip=true, trim= 0 0 0 50]{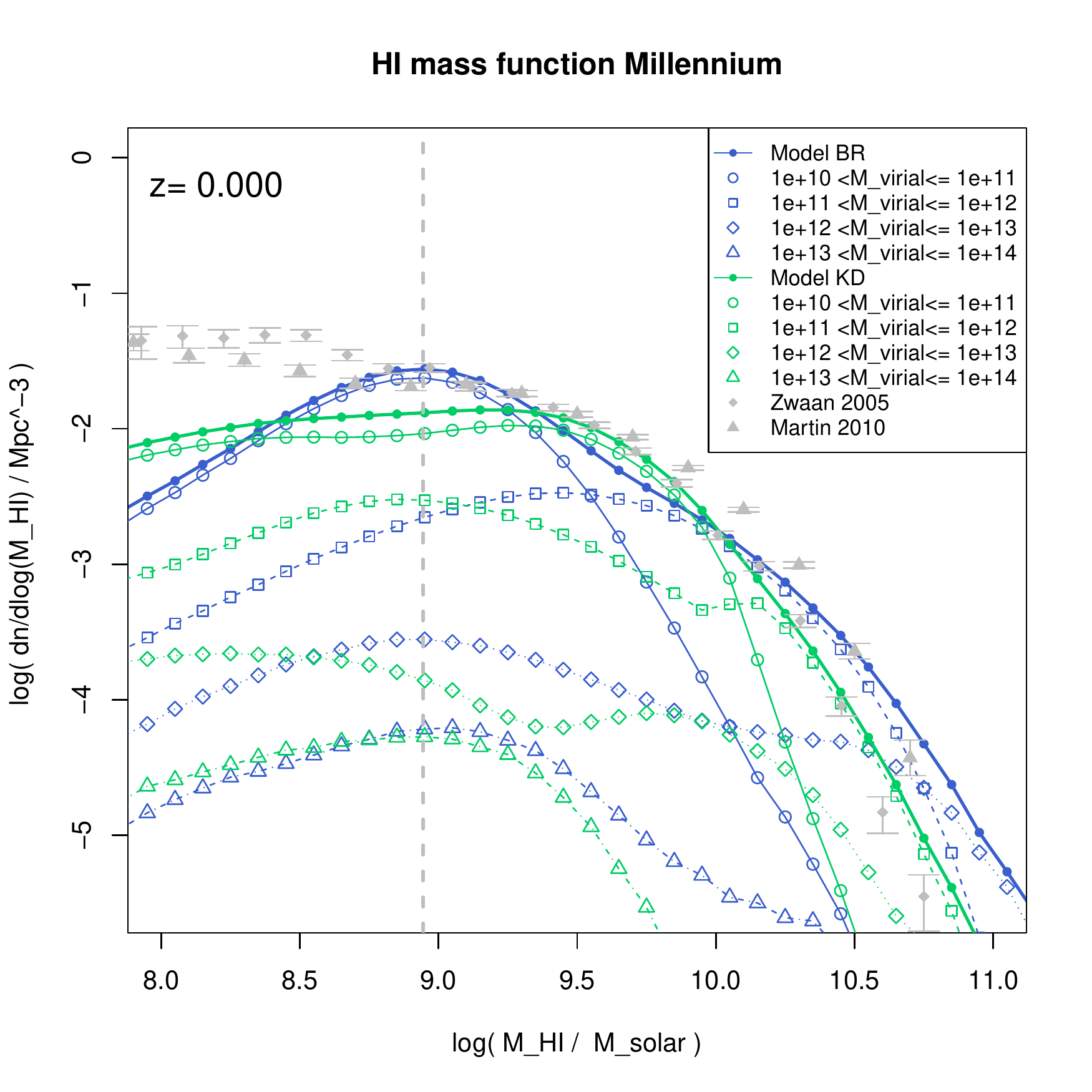}\\
\includegraphics[scale=0.5, clip=true, trim= 0 0 0 50]{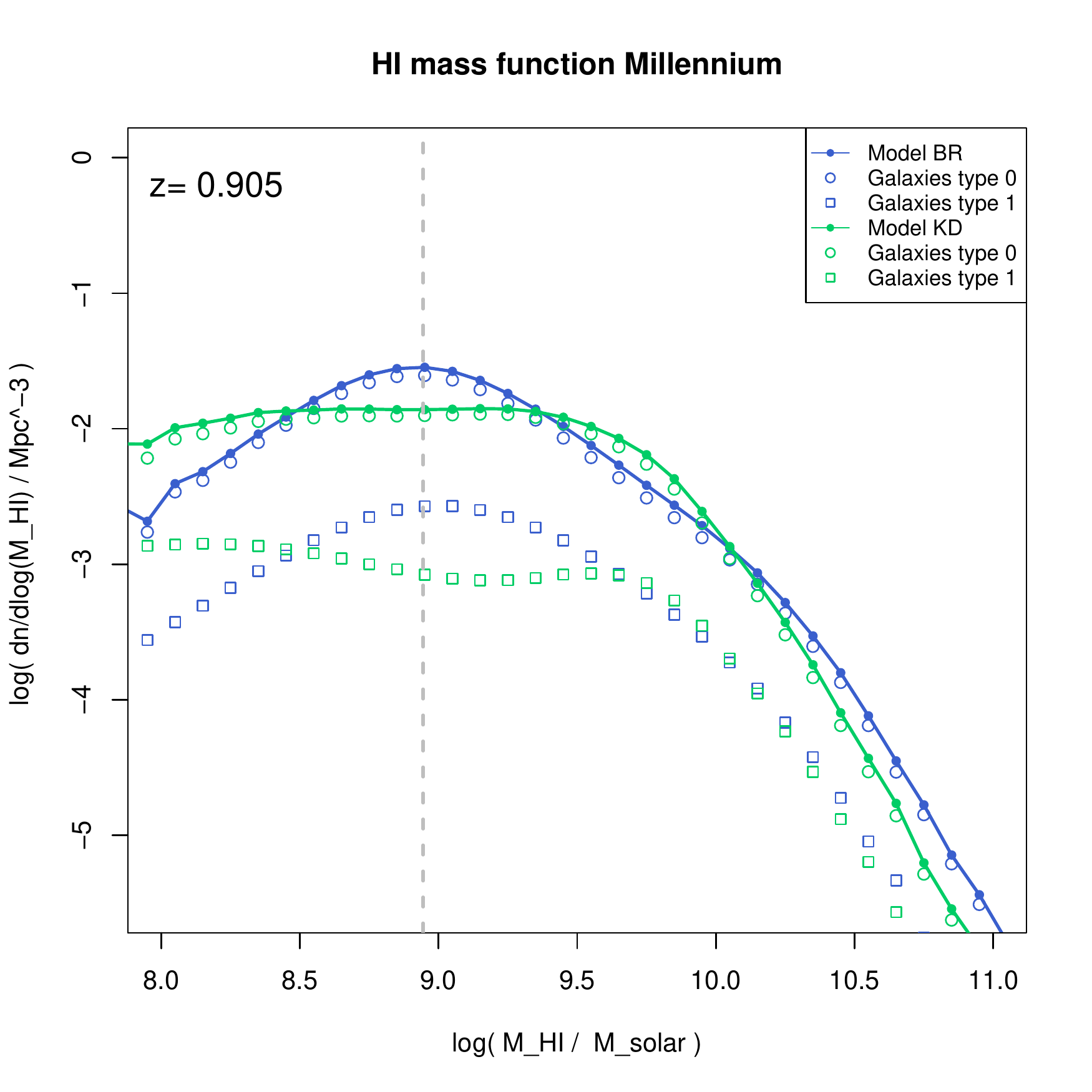}
\includegraphics[scale=0.5, clip=true, trim= 0 0 0 50]{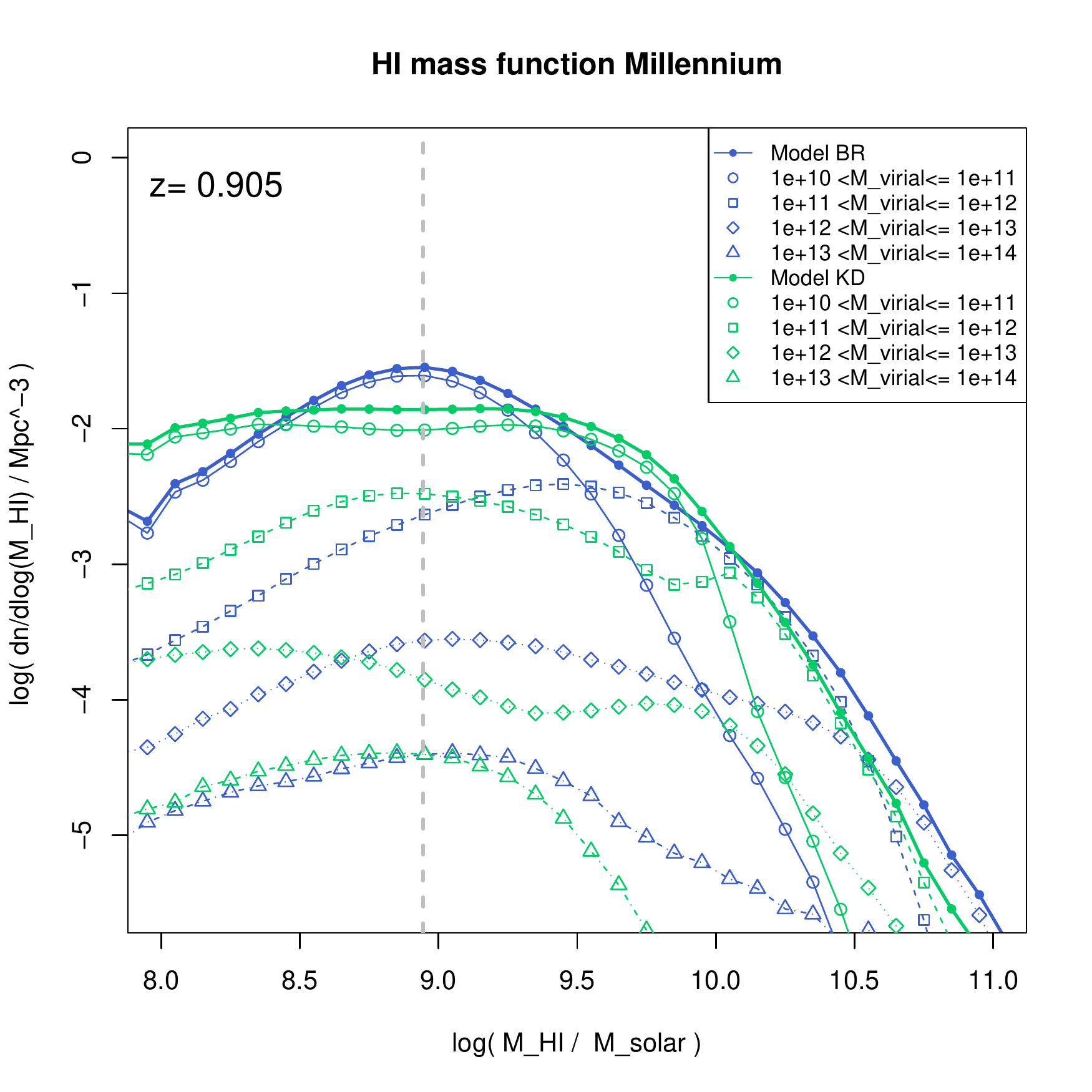}
\end{tabular}
\caption[HI mass function]{HI mass function for different star formation models. The upper row shows the HIMF for $z=0.0$ in comparison to measurements by HIPASS \citep{2003AJ....125.2842Z} and ALFALFA \citep{2010ApJ...723.1359M}. The models are in reasonable agreement with the data up to the particle mass resolution limit at $8.8\times10^8$\Msun which is indicated by the dashed grey line. The lower row shows the HIMF for $z=0.905$, there is only moderate redshift evolution seen in the HIMF, which is slightly stronger for the BR model. The panels in the left column show the HIMF split into the contributions by centrals (type 0)  and satellite galaxies (type 1).  The HIMF in the right panels are split according to the virial mass of their host halos. We can see that the lightest halos contribute largest HI masses for both models, however the high mass end is dominated by bigger halos $(10^{12}-10^{13})$\Msun in the BR model.}
\label{HIMF}
\end{figure*}
Fig.~\ref{HIMF} displays the HI mass function (HIMF) of the two different star formation recipes for $z=0$ in the upper row and $z=0.905$ in the lower row. In the first column, we have split the HIMF into the contributions from central (type 0) and satellite galaxies (type 1) and in the second column we have binned the HIMF according to the virial masses of the host halo of the galaxy. 

Firstly, in the left top panel, we compare the models given as the coloured solid lines with the observational data from HIPASS \citep{2003AJ....125.2842Z} and ALFALFA \citep{2010ApJ...723.1359M} marked with grey symbols. We would like to highlight how well the simulations fit the data down to HI masses of $10^9$\Msun~given that the SAGE model is not tuned to produce the HIMF at redshift zero. We believe both models reasonably well predict the HIMF given the uncertainties in the theoretical descriptions for star formation and its connection to the atomic phase of cold gas. Given the mass resolution of the simulation, we can not make any reliable predictions for HI masses smaller than $10^9$\Msun. It has been suggested that a large fraction of the atomic hydrogen is located in the low mass end of the HIMF \citep{2013MNRAS.428.3366K,Kim:2015ht} which is unresolved in our current simulation resolution. 
This could lead to an underestimation of the HI power spectrum on all scales, however, we do not expect this effect to alter the presented results on the cross-correlation.

In the upper left panel, the central galaxies predominantly contain atomic gas over the whole range of HI masses for all models. The model including AGN feedback shows very similar behaviour as the KD model and is left out the remaining plots for simplicity.
The second panel in the upper row illustrates the contribution of the halos. For the KD model, the high mass end is dominated by mid-mass halos with $10^{11}\Ms <M_{\rm vir}<10^{12}\Ms$. From HI masses smaller than $10^{9.5}$\Msun, the majority of the HI is located in the lightest halos with $10^{10}\Ms <M_{\rm vir}<10^{11}\Ms$. The BR model predicts similar behaviour. However, the very high mass end, which is slightly over predicted compared to observations in this model, is caused by galaxies in halos with $10^{12}\Ms <M_{\rm vir}<10^{13}\Ms$.

In the second row, the HIMF for $z=0.905$ exhibits the same characteristics as shown in figures for $z=0.0$. A moderate redshift evolution can be seen in comparison to the upper panels. There are no tight observational constraints for redshifts beyond the local Universe due to the weakness of the HI signal and poor constraints for the low mass end of the HIMF. It is a key objective of future radio surveys (e.g. FAST, ASKAP and SKA) to improve the  observations of the HIMF at high redshift.

\cite{2011MNRAS.416.1566L} previously implemented a neutral hydrogen model in their SAM using models similar to the ones presented in Sec.~\ref{SFmodels}, and studied the evolution of the gas contents at higher redshift in \cite{2011MNRAS.418.1649L}. They favoured the BR model over the KD star formation recipe when using the variant of the SAM chosen as the fiducial model for their implementation. For $z=0$, \cite{2011MNRAS.418.1649L} found similar distributions of the HI content with virial mass, and that satellite galaxy predominately contribute to the HIMF on masses smaller than $10^{7.5}\Ms$ which is beyond our resolution limit. The redshift evolution of the HIMF is similarly weak between redshift 0 and 1 as in our simulation.
\subsection{Colour Diagrams}
\begin{figure*}
\includegraphics[width=0.33\textwidth, angle=0, clip=true, trim= 0 0 0 0]{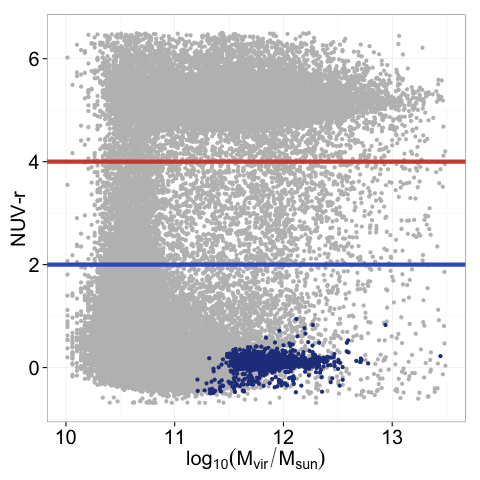}
\includegraphics[width=0.33\textwidth, angle=0, clip=true, trim= 0 0 0 0]{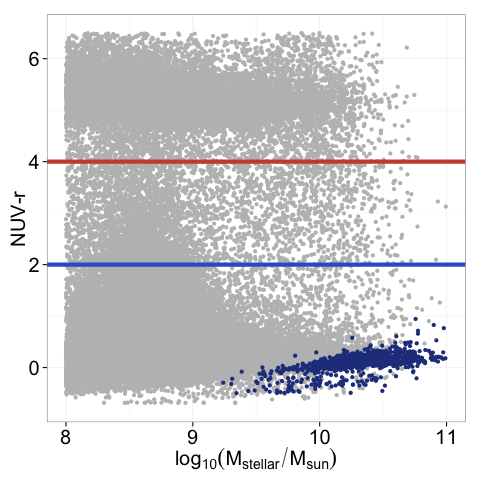}
\includegraphics[width=0.33\textwidth, angle=0, clip=true, trim= 0 0 0 0]{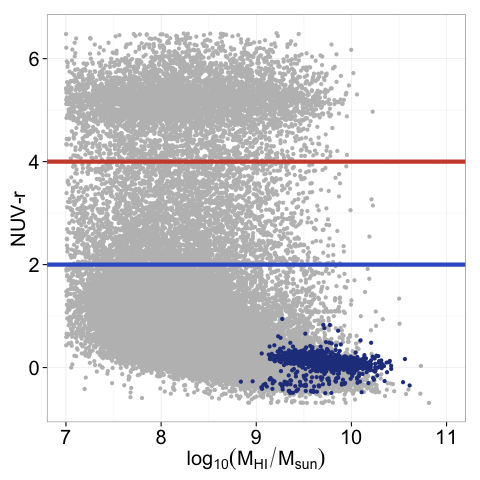}\\

\caption[]{(NUV-r) colours depicted for the KD model as a function  virial halo mass (left), stellar mass (middle) and HI mass (right) for $z=0.905$. We show the full simulation as the grey scatter and indicate the division into the red and blue galaxies by the coloured lines. The case of the WiggleZ selected galaxies are marked as dark blue circles on top. 
 }
\label{ColoursKD}
\end{figure*}
\begin{figure}
\centering
\includegraphics[width=0.33\textwidth, angle=0, clip=true, trim= 0 0 0 0]{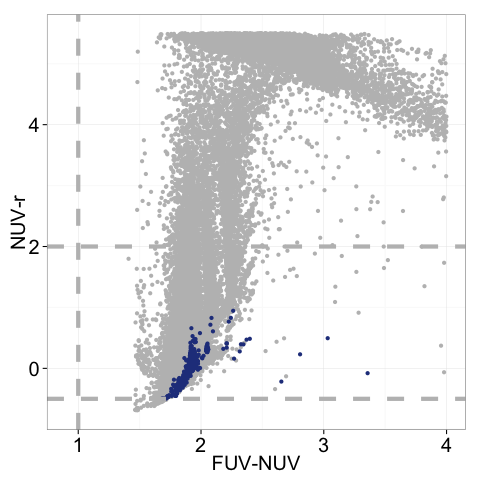}
\caption[]{(NUV-r) colour depicted for the KD model as a function of (FUV-NUV) with grey dashed lines which denote the WiggleZ selection cuts. The WiggleZ selected galaxies are marked as dark blue circles on top. The dominant selection effect is the selection of the brightest galaxies in the NUV band with $m_{\rm NUV} < 22.8$. 
 }
\label{ColoursWig}
\end{figure}

In Fig.~\ref{ColoursKD}, we present scatterplots of the simulation with star formation model KD in relation to the galaxy colour, quantified by the colour $(\rm NUV-r)$ used as a proxy for SF history. The quiescent galaxy population is located above the red horizontal line and the star-forming one below the blue horizontal line.  We have used 1\% of the simulated galaxies for the graphics, such that the number of points do not represent the complete galaxy numbers.
We show the relation of the colour to the galaxy properties such as the virial mass, the stellar mass and the atomic hydrogen mass. We see that the quiescent galaxies span a wider range of virial masses occupying the most massive halos. Furthermore, the HI abundance is in general lower in the quiescent population such that parts are stripped of their cold gas. However, the remaining red galaxies represent the whole HI mass spectrum. For  actively star-forming galaxies, the galaxy colour and the HI mass are weakly correlated, such that the highest HI abundances connect to the bluest galaxies, indicating high star-formation activity.
\begin{itemize}
\item{WiggleZ case \\
Fig.~\ref{ColoursWig} pictures the WiggleZ selection criteria as given in Equ.~\ref{Wigsel}, where the grey dashed horizontal lines represent the colour selection and the dashed vertical line the Lyman-break-galaxies criteria. For our cube with galaxies at $z=0.905$ the Lyman-alpha break redshift selection criteria is satisfied by all galaxies as expected. The WiggleZ selected galaxies are present at the bottom end of the blue galaxies which is caused by the magnitude cut in NUV, such that WiggleZ only selects UV bright galaxies for this redshift. 

From Fig.~\ref{ColoursKD} we can infer that the WiggleZ criteria select galaxies hosted in medium sized halos with $M_{vir}\sim10^{12}$\Msun~with very high stellar mass $M_*\geq10^{10}$\Msun~and additionally very high HI mass with $M_{\rm HI}\geq10^{9}$\Msun. This implies that the WiggleZ selected galaxies are an extreme subset of the overall blue, star-forming galaxy population. We find similar relations in the scatterplots for the other two simulation models. We study the selection effects of the WiggleZ cuts and their relation to the HI mass of the galaxies further by considering their HI scaling relation in the following section.
}
\end{itemize}
\subsection{HI Scaling Relations}
Observed HI scaling relations have been presented for low redshift galaxies with stellar masses greater than $10^{10}\Ms$ in \cite{Catinella:2009eo}, these relations show that there is a strong anti-correlation between the fractional HI mass and galaxy colour $(\rm NUV-r)$. In a different project, \cite{Cortese:2011sj} confirmed previous results and investigated the connection of the scaling relations to the galaxy environment. They found that HI-poor galaxies are more likely to be found in galaxy clusters than in low-density environments. In the following, we present the scaling relations of our simulation to illustrate the correlation between HI gas properties and galaxy colour which are used as selection criteria for the intensity maps and the galaxy populations.

\label{sec_scaling}
\begin{figure*}
\subfigure[KD simulation]{
\includegraphics[width=0.33\textwidth, angle=0, clip=true, trim= 0 0 0 0]{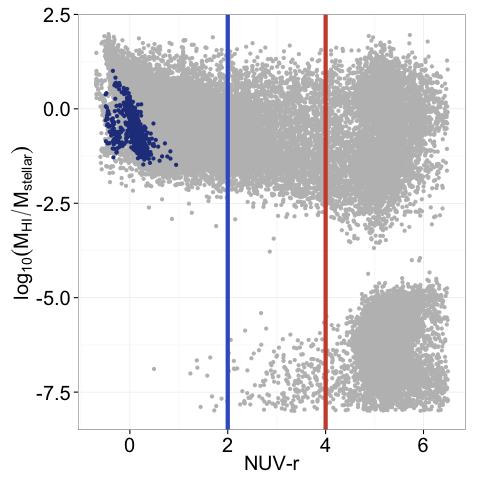}
}\subfigure[KD AGN simulation]{
\includegraphics[width=0.33\textwidth, angle=0, clip=true, trim= 0 0 0 0]{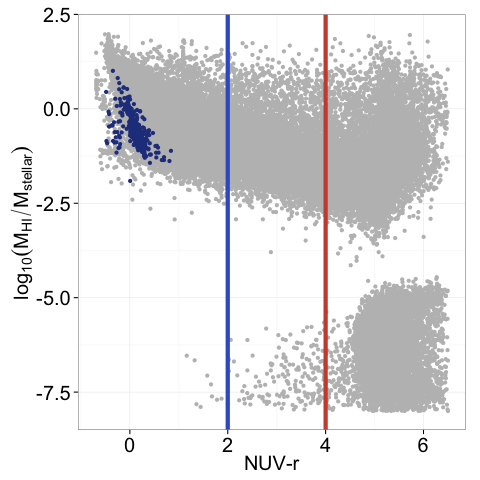}
\label{HI_colour_KD_AGN}
}\subfigure[BR simulation]{
\includegraphics[width=0.33\textwidth, angle=0, clip=true, trim= 0 0 0 0]{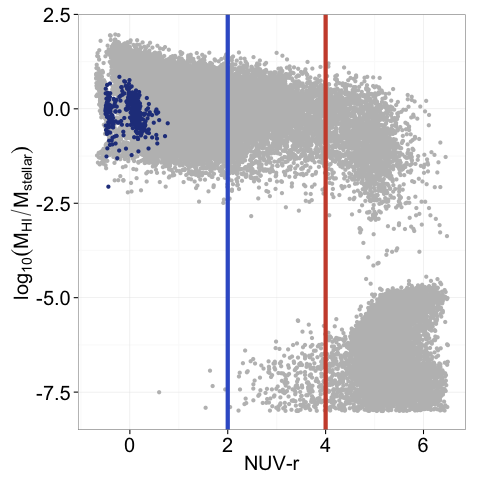}
}
\caption[HI scaling relations]{The HI scaling relation is given as the fraction of HI mass to stellar mass as a function of galaxy colour for different SF models at $z=0.905$. All panels show the full simulation as grey density fields. The dashed lines indicate colour cuts between blue ($(\rm NUV-r)<2$), green ($2<(\rm NUV-r)<4$) and red galaxies ($4 <(\rm NUV-r)$). We can see that the WiggleZ selected galaxies have a tight scaling relation and tend to have a very high fraction of HI mass.}
\label{HI_colour}
\end{figure*}

In Fig.~\ref{HI_colour}, we depict the fraction of HI mass to stellar mass as a function of galaxy colour for the three different models, again only using 1\% of the simulated galaxies for computational ease. The coloured lines mark the different galaxy colour domains as blue ($(\rm NUV-r)<2$) and red galaxies ($4 <(\rm NUV-r)$). The underlying grey scatter of the full galaxy population shows how the blue population is in general HI rich compared to their stellar mass and, in many cases, the galaxy mass is dominated by neutral hydrogen. The red galaxies are divided into two different regimes, high fractions of HI gas and very low fractional HI density. 
Fig.~\ref{HI_colour_KD_AGN} illustrates how AGN feedback prevents star-formation by heating the gas and reducing the amount of  HI in the galaxies,  lifting the division between blue and red galaxies. We further note that the WiggleZ cut (marked by dark blue circles), while selecting extremely HI rich galaxies, does not favour galaxies with the highest fraction of HI gas to stellar mass for all SF recipes.
\begin{figure}
\subfigure[Colour selection]{
\includegraphics[width=0.5\textwidth, angle=0, clip=true, trim= 0 0 0 0]{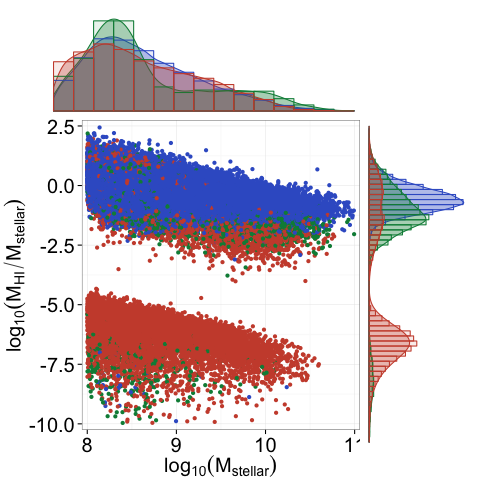}
\label{HI_stellar_a}
}
\subfigure[WiggleZ selection]{
\includegraphics[width=0.45\textwidth, angle=0, clip=true, trim= 0 0 0 0]{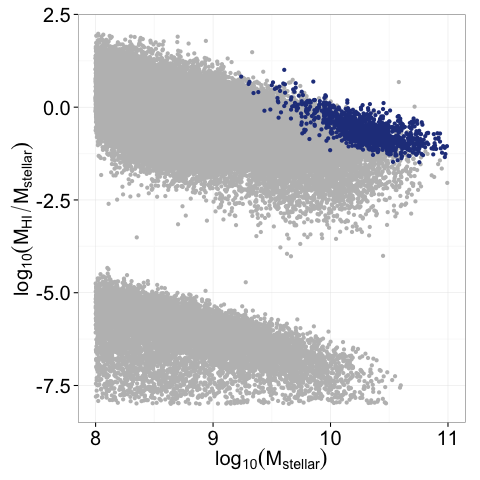}
\label{HI_stellar_b}
}
\caption[HI scaling relations]{The HI scaling relation is given as the fraction of HI mass to stellar mass as a function of stellar mass for the KD SF model at $z=0.905$ for 1\% of the simulated galaxies. The upper panel illustrates the colour selected galaxies, blue, green and red, as a scatter plot. The lower panel shows the full simulation as grey scatter and the WiggleZ selected blue dots. }
\label{HI_stellar}
\end{figure}

In Fig.~\ref{HI_stellar}, we illustrate the HI scaling relation of the HI mass over stellar mass as a function of stellar mass. We show an example using the KD star formation recipe since all models exhibit similar scalings. 
In the upper panel we show the relation between $M_{\rm HI}/M_*$ and $M_*$ for the galaxies divided by their colours with the cuts shown in Fig.~\ref{HI_colour}, where we have added a population of green galaxies with ($2 <(\rm NUV-r)<4$). As expected, the star-forming galaxies dominate the high HI mass regime. However, the figure also illustrates that quiescent galaxies contribute considerably to the fraction of galaxies with high relative HI masses, leading to a bimodal character of their scaling relation.
In the lower panel of Fig.~\ref{HI_stellar_b}, we see how the WiggleZ galaxies, as the population with the highest apparent NUV magnitudes, are given by the galaxies with the highest stellar mass with a relatively high fraction of HI to stellar mass. This population is therefore a very unique subset of the blue galaxies, probing only a small parameter space of star-forming galaxies.

\section{Power Spectrum Results and Discussion}
\label{sec4}
In the following section, we present the power spectra of the intensity maps, the colour-selected galaxies, the WiggleZ-selected galaxies and the cross-correlation of intensity maps and galaxies.

We compute the spectra using a 3-dimensional cube of galaxy over densities given as $\delta_i=(N_i - \bar N_g)/ \bar N_g$ for each pixel $i$ where $N_i$ is the number of galaxies per pixel and $\bar N_g$ is the mean galaxy density. We Fourier-Transform the over density cube into $ \tilde \delta(\vec k_i)$ and spherically average over wavenumbers $\vec k_i$ fulfilling the condition $(k-\Delta k) \leq |\vec k_i | < (k+\Delta k)$ to compute the power spectrum $P_g(k)=<|\tilde\delta(k)|^2>$ in units of $(\Mpch)^3$. We convert this into the dimensionless spectrum $\Delta^2_g(k) = k^3P(k)/(2\pi^2)$.

We use a similar description for the intensity maps which are given as temperature fluctuations $\delta_T=(T_i - \bar T)$ in units of mK. The resulting power spectra  $\Delta^2_T(k)$ are given in units of $\rm mK^2$ and  the cross power spectra of galaxies and intensity maps $\Delta^2_X(k)$ in units of mK.

The following power spectrum figures present the dimensionless spectrum $\Delta^2(k)$ in the upper panel and the relative scale-dependent shape of the spectra by dividing out the reference model and the fitted scale-independent bias $\tilde b$ in the lower panel. We estimate the scale-independent bias for each via a maximum likelihood fit to a theoretical reference model on large scales $0.08\hM<k<0.2\hM$. For the reference model we use a linear prediction of the power spectrum based on the cosmology used for the Millennium simulation given as $(h=0.73, \Omega _{\rm m}=0.25, \Omega _{\rm b}=0.045, n_{\rm s}=1.0, \sigma_{8}=0.9)$.

\subsection{Galaxy Power Spectrum}
\begin{figure}
\includegraphics[width=0.5\textwidth, clip=true, trim= 0 0 0 50]{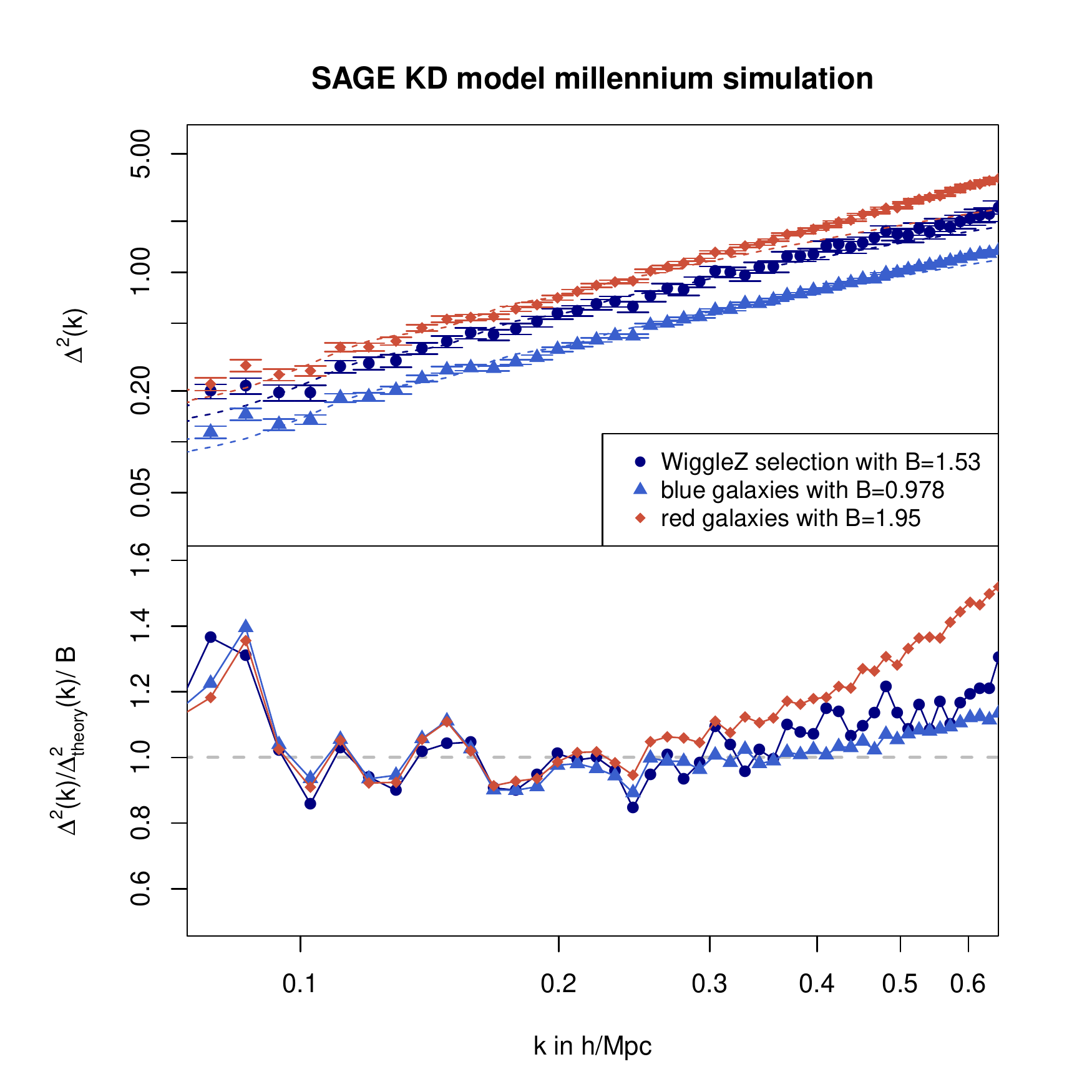}
\caption{Galaxy power spectrum of the KD model for different types of galaxies: WiggleZ selected, blue and red. The lower panel shows the scale-dependent dimensionless clustering as a function of $k$. Red galaxies are more strongly clustered, and have a stronger scale-dependence of clustering as expected from observations and simulations. Note that the factor $B$ given in the legend converts as $B=\tilde b_{\rm g}^2$.}
\label{PS_gal}
\end{figure}

In Fig.~\ref{PS_gal}, we show the power for the KD model divided into red and blue galaxies as well as WiggleZ selected galaxies as described in Sec.~\ref{sec_scaling}. The galaxy power spectrum is connected to the underlying dark matter power spectrum via $\Delta^2_{\rm g}(k)=b_{\rm g}^2 (k) \Delta^2_{\rm dm}(k)$. We remove the Poisson noise contribution from the presented power spectra which is given as the inverse of the mean galaxy density such that we subtract the term $\Delta^2_N=k^3 (\bar N_{\rm g})^{-1}/(2\pi^2)$.

As expected, we find that red galaxies show stronger clustering on scales smaller than $k\approx0.3\hM$ which is in agreement with observations (e.g. \citealt{Guzzo:1997ke,Norberg:2001fi,Heinis:2009cs,2008MNRAS.385.1635S}). The estimated scale-independent bias of quiescent galaxies is higher than for blue galaxies. We find that the other star formation models exhibit a similar trend as the KD model, and that the broadband clustering of each galaxy population does not depend critically on the star-formation model. The scale-independent biases $\tilde b_{\rm g}$ estimated on large scales $k<0.2\hM$ for each model and galaxy cut are given in Tab.~\ref{table_b}.

\begin{figure}
\includegraphics[width=0.5\textwidth, clip=true, trim= 0 0 0 50]{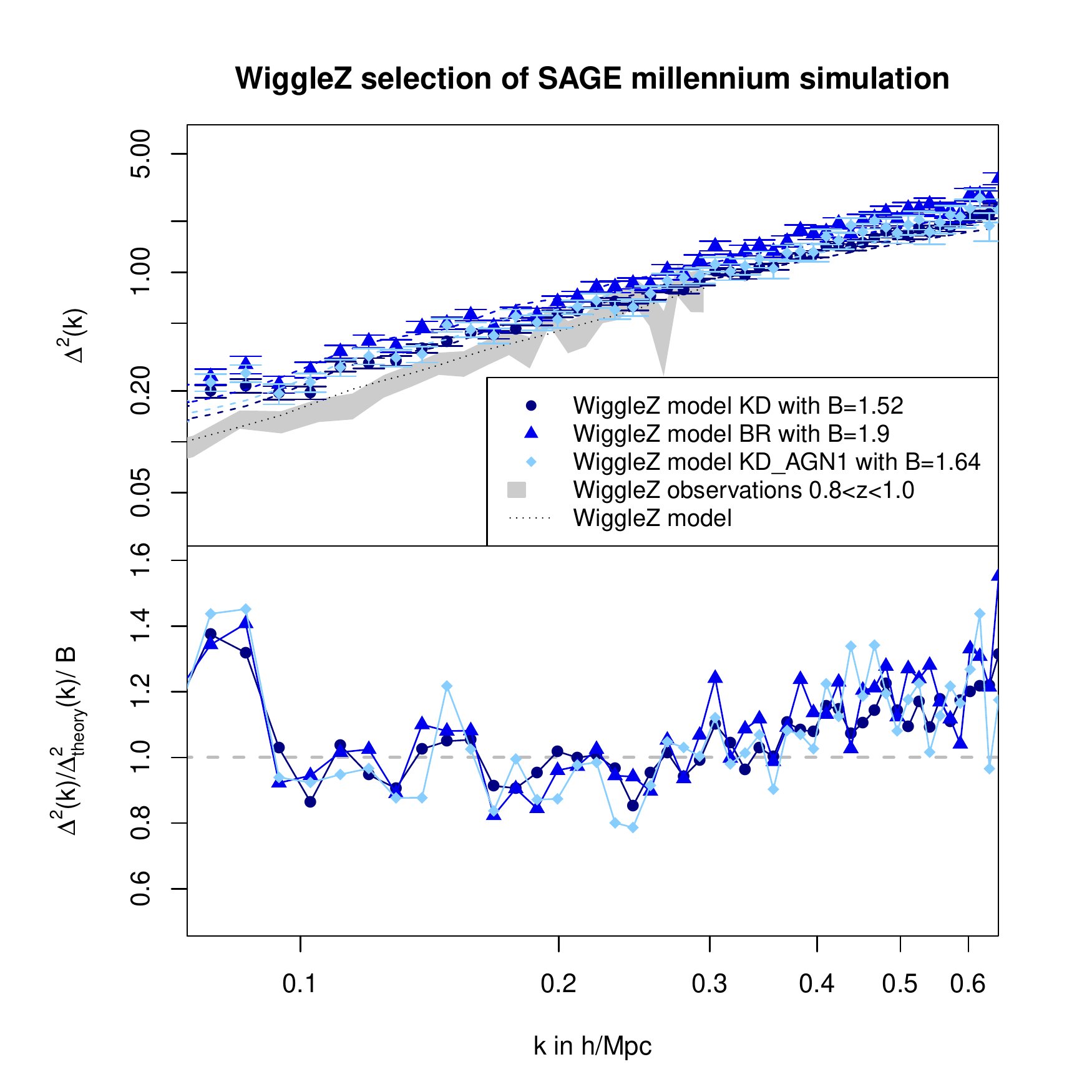}
\caption{Galaxy power spectrum of WiggleZ selected galaxies simulated using different SF models. The scale-independent bias of the models differs slightly due to different star formation mechanism which influences the number of UV bright galaxies, but the overall shape of the power spectrum agrees reasonable well for all models. Note that the factor $B$ given in the legend converts as $B=\tilde b_{\rm Wig}^2$.}
\label{PS_Wig}
\end{figure}

In Fig.~\ref{PS_Wig}, we show power spectra for the WiggleZ mock catalogues in comparison with the observations for galaxies with $0.8<z<1.0$. The significant errors on the redshift estimates for the high redshift end of the WiggleZ survey, and the steep decrease of the WiggleZ redshift distribution bias the power spectrum towards $z\approx0.8$. This partly causes the difference of the amplitude of the models and the data, in addition to the use of different underlying cosmological models. The Millennium simulation was computed with $\sigma_8=0.9$ which is an over-estimate according latest measurements. The three models produce similar galaxy biases with values of $1.19 < \tilde b_{\rm Wig}< 1.32$ which are comparable to the measured bias of $1.20 \pm0.06$ given in \cite{Blake:2010uf}. Given the considerable errors on the WiggleZ observations in this high redshift regime, this agreement is sufficient for our theoretical study which does not aim to simulate any systematic effects. 

We note that the different star-formation recipes do not influence the clustering shape shown in the lower panel of Fig.~\ref{PS_Wig}. However, the number density of WiggleZ-selected galaxies which are bright enough for the UV selection is sensitive to the star formation model. The KD model predicts a galaxy density of $\sim 10 \times10^{-4} (h\rm{Mpc}^{-1})^{3}$ whereas  AGN feedback reduces the density to $3\times10^{-4}(h\rm{Mpc}^{-1})^{3}$. The BR star formation recipe predicts a lower density with $4\times10^{-4}(h\rm{Mpc}^{-1})^{3}$. The observational number density around $z\approx 0.9$ is lower with values around $0.5\times10^{-4} (h\rm{Mpc}^{-1})^{3}$ which can be explained by incompleteness and the low redshift success rate such that the number of sources is significantly reduced.

\subsection{Intensity Mapping Power Spectrum}
\begin{figure}
\includegraphics[width=0.5\textwidth, clip=true, trim= 0 0 0 50]{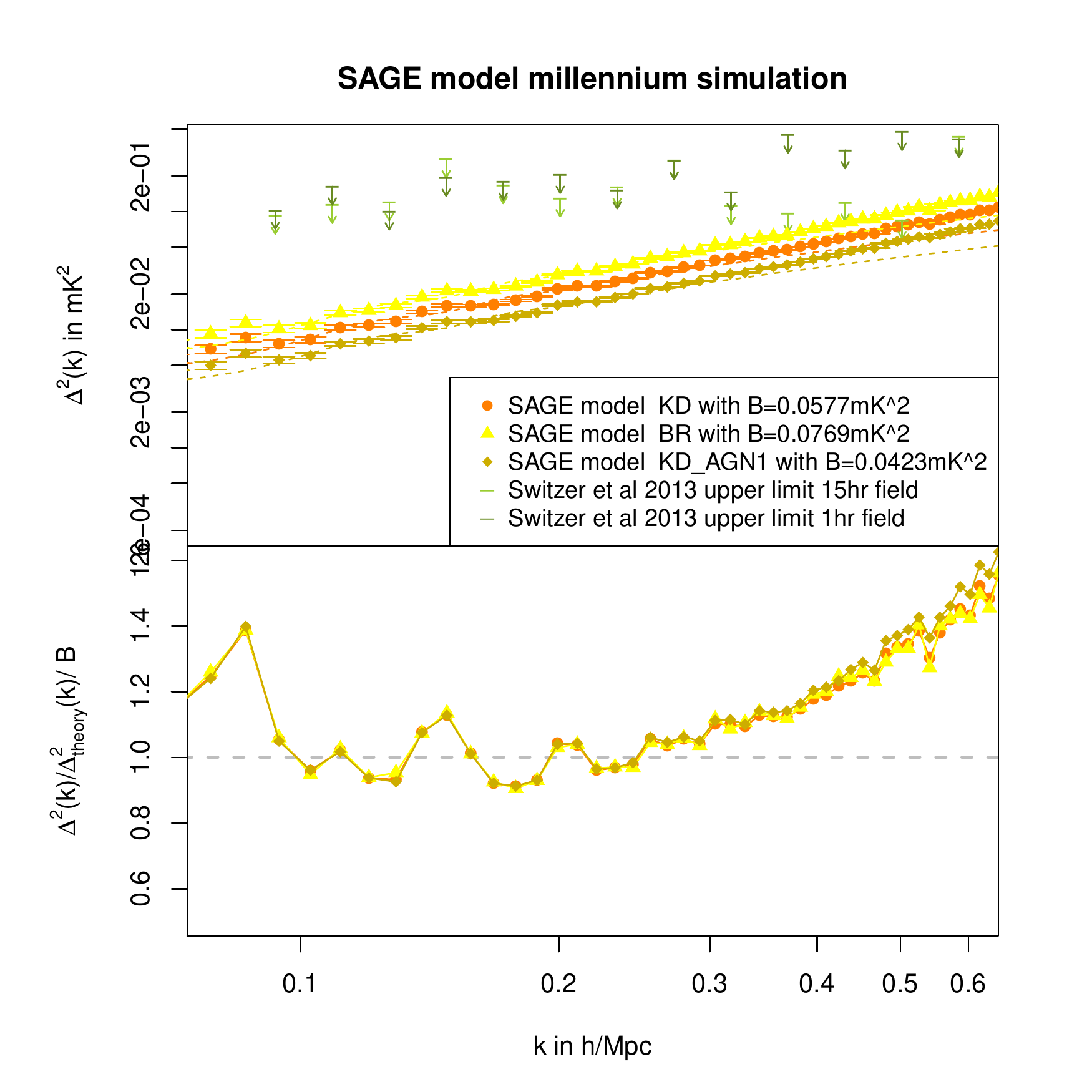}
\caption{Intensity Mapping temperature power spectrum for different SF models in units of $\rm mK^2$ at $z=0.905$ in comparison to the upper limit measurements of the GBT team (\citealt{2013MNRAS.434L..46S}). We can see that different models produce different broadband amplitudes due to different \OHI~given by the models. Lower panel shows the dimensionless clustering shape of the power spectrum and we see that models reasonable agree. Note that the factor $B$ given in the legend converts as $B=\tilde b_{\rm HI}^2 \bar T_{\rm HI}^2$.}
\label{PS_IM}
\end{figure}

The power spectrum of the intensity maps corrected for the telescope beam in units of $\rm mK^2$ is presented in Fig.~\ref{PS_IM} where we compare the different recipes for star formation. The intensity mapping power spectrum is related to the underlying dark matter distribution via $\Delta^2_T(k)=\bar T_{\rm HI}^2 b_{\rm HI}^2 (k) \Delta^2_{\rm dm}(k)$. We can use a model to express the HI temperature as a function of redshift and cosmological energy densities of matter and HI such that 
\begin{equation}
\bar T_{\rm HI}=317\rm{mK} \times \Omega_{\rm HI}; \textrm{ for } z=0.905.
\label{THI}
\end{equation}
The HI bias and \OHI~determine the amplitude of the power in a degenerate fashion, where any scale-dependency is caused by non-uniform distribution of HI in galaxies, i.e. dark matter halos. 

The models predict different values for the amplitude coefficients of the power spectrum, with values of approximately $0.7 \times10^{-3}$, which is within the limits of the observational constraints of \OHI$\tilde b_{\rm HI}=[0.62\pm 0.23]\times 10^{-3}$ (\citealt{2013MNRAS.434L..46S}) for median redshift of 0.8. The exact values for \OHI$\tilde b_{\rm HI}$ predicted by each model are listed in Tab.~\ref{table_b}. All models predict the same scale-dependent clustering shape as seen in the lower panel which establishes a robust model for the intensity maps. 

We show the upper limit measurements for the auto-power spectrum of the GBT data (\citealt{2013MNRAS.434L..46S}) in Fig.~\ref{PS_IM} for comparison with our simulations. These constraints on the intensity power spectrum are two orders of magnitude higher than the theoretical predictions, mainly caused by instrumental systematics. This encourages the cross-correlation of intensity mapping data with galaxy surveys.
\begin{itemize}

\item{\textbf{Intensity Mapping Poisson Noise}}\\
\label{IMSN}
By construction, the intensity mapping power spectrum contains a Poisson noise contribution from the sampling of the HI masses from individual galaxies. The so-called shot noise term for galaxy power spectra can be approximated as the inverse of the observed galaxy density. For intensity mapping power spectra, the contribution of each galaxy is weighted by the individual HI flux, the term is convolved by the beam and multiplied by the mean HI temperature. In difference to a galaxy redshift survey, for intensity mapping observations, every single object containing atomic hydrogen contributes to the maps such that the shot noise is the minimal possible and independent of experimental set-up. It is possible to subtract the shot noise for a given simulation, however we choose not to subtract any contribution since it is not feasible to do such with observations where the HI galaxy density is unknown.  
\end{itemize}
\subsection{Cross-Correlation}
\begin{table*}
\begin{tabular}{ c | c | c |c | c | c | c | c}
Model & $\tilde b_{\rm WiggleZ}$ & $\tilde b_{\rm blue}$ & $\tilde b_{\rm red}$  & \OHI$\tilde b_{\rm HI}$ & $\tilde r_{\rm WiggleZ}$ &$\tilde r_{\rm blue}$&$\tilde r_{\rm red}$\\
  \hline			
 KD& 1.23 & 0.98 & 1.39 & $7.58\times 10^{-4}$ & 1.01 & 0.98 & 0.96 \\
 KD AGN& 1.28 & 0.94 & 1.30 & $6.49\times 10^{-4}$ & 1.01  & 0.98 & 0.97 \\
 BR &   1.37 & 1.05 & 1.33 & $8.75\times 10^{-4}$ &1.04 & 0.99 & 0.98 \\
  \hline  
\end{tabular}
\caption{The scale-independent galaxy bias $\tilde b$ for the different galaxy populations (WiggleZ selection, blue and red), HI bias times the HI energy density and the respective scale-independent cross-correlation coefficients for the three different star-formation models estimated over wavenumber $0.08\hM<k<0.2\hM$.}
  \label{table_b}
\end{table*}
\begin{figure}
\includegraphics[width=0.5\textwidth, clip=true, trim= 0 0 0 50]{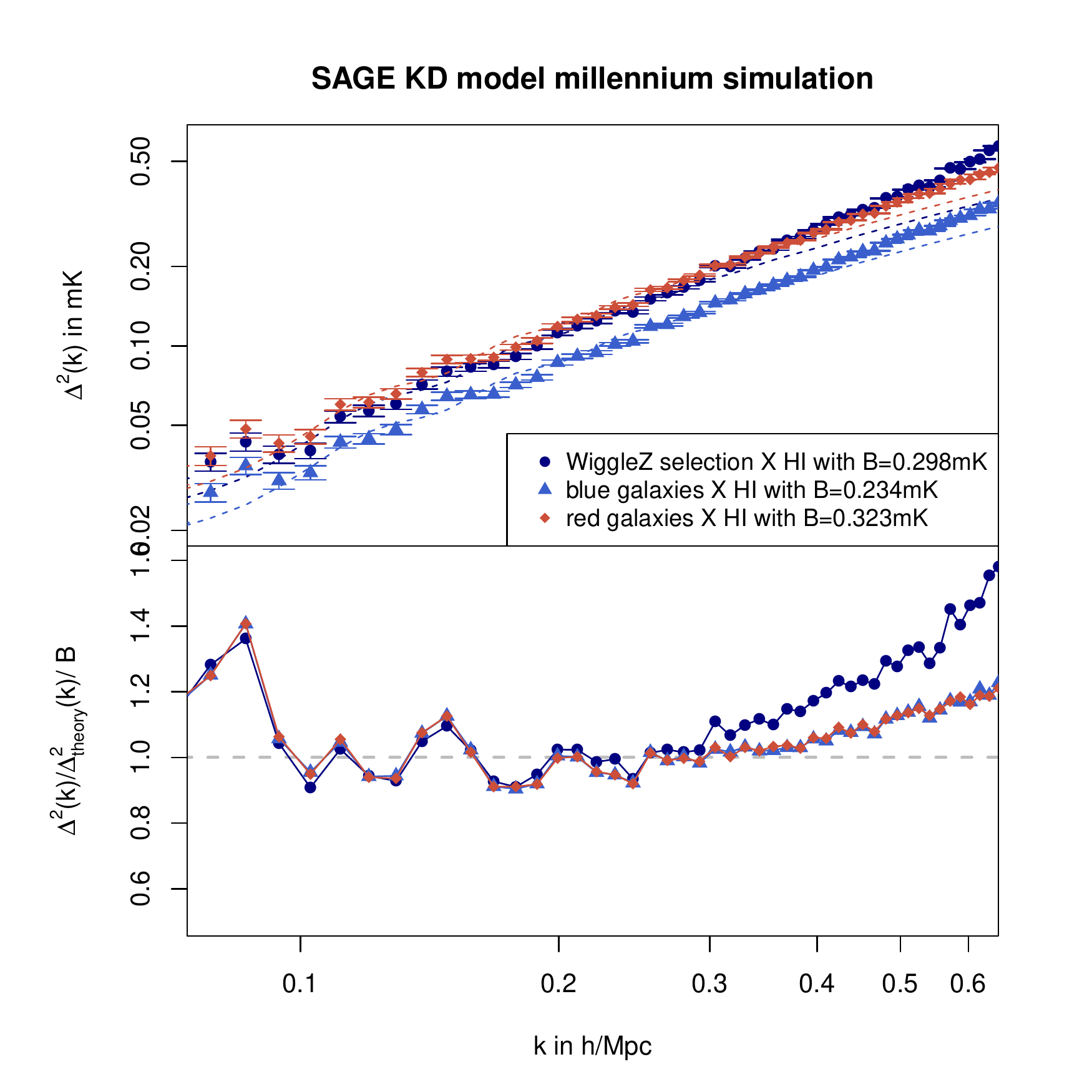}
\caption{Cross power spectrum of galaxies and intensity maps given by the KD model in units of mK in the upper panel. The lower panel shows the dimensionless clustering behaviour of the cross-correlation demonstrating how the high clustering of WiggleZ galaxies with HI on scales smaller than $k\approx0.3$. Note that the factor $B$ given in the legend converts as $B=\bar T_{\rm HI} \tilde b_{\rm HI} \tilde b_{\rm g}$.}
\label{PS_X}
\end{figure}

\begin{figure}
\subfigure[SAGE model HI description]{
\includegraphics[width=0.45\textwidth]{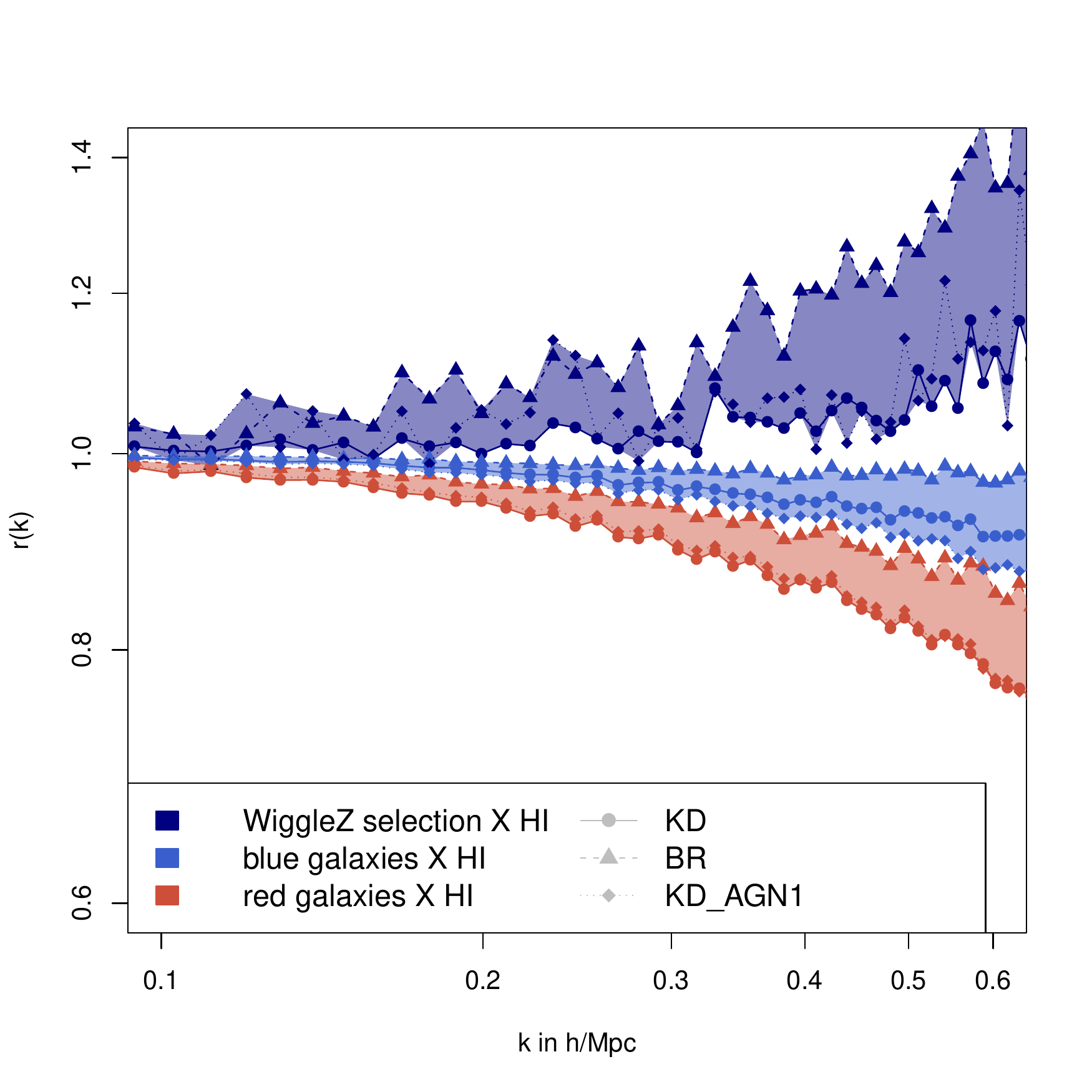}
\label{r_k_model}
}
\subfigure[Random HI clustering]{
\includegraphics[width=0.45\textwidth]{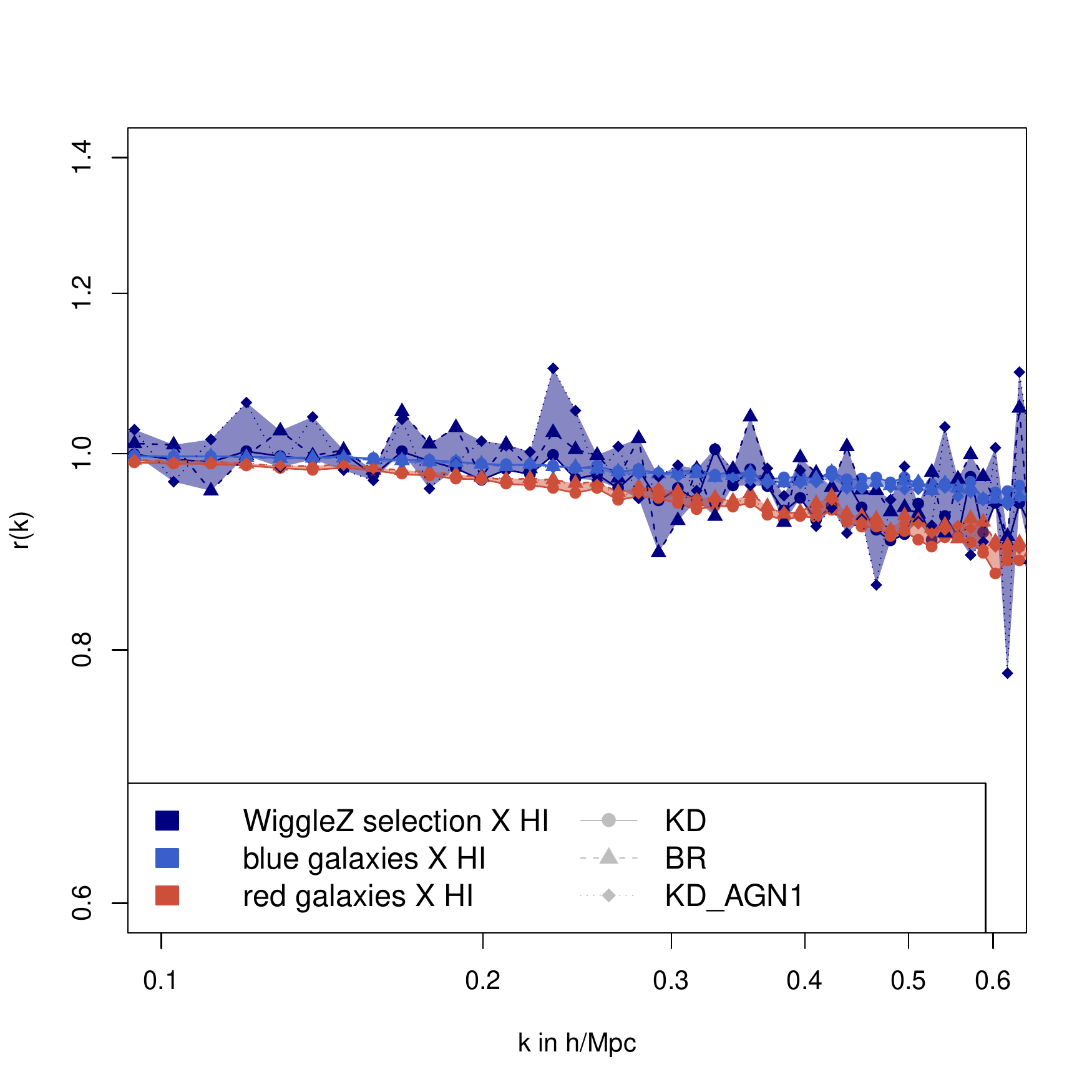}
\label{r_k_random}

}
\caption{Correlation coefficients $r(k)$ of the intensity maps with quiescent (red), star-forming (light blue) and WiggleZ-selected (dark blue) galaxy populations for different star-formation recipes. The shaded areas mark the range of the star formation recipes for each galaxy population. The lower panel presents a null-test with randomly distributed HI masses.}
\label{r_k}
\end{figure}

In Fig.~\ref{PS_X}, the cross-correlation power spectrum of the intensity maps with the different galaxy populations is presented. The cross-correlation power spectrum is calculated as
\begin{equation} 
\Delta^2_{\rm X}(k)=\bar T_{\rm HI} b_{\rm HI}(k)  b_{\rm g}(k) r(k) \Delta^2_{\rm dm}(k)
\end{equation}
where $r(k)$ is the cross-correlation coefficient of the two probes.
In the following, we present constraints for \ObHI~ using the temperature conversion presented in Equ.~\ref{THI}. 
We choose to fit a scale-independent amplitude of the correlation using $\tilde b$ for $k<0.2\hM$. 

The cross-power is shown in Fig.~\ref{PS_X} using the KD model for the different galaxy populations. We find that the scale-dependent clustering on small scales $k>0.3\hM$ is indistinguishable for quiescent and star-forming galaxies. 
The power on small scales is determined by the correlation between HI content and the colour of the galaxy, which is not a linear relation as shown in Fig.~\ref{ColoursKD}. We find that many quiescent galaxies have high HI contents and thus positively correlate with intensity maps dominated by HI rich galaxies. 

On the other hand, WiggleZ-selected galaxies strongly cluster with the intensity maps on scales smaller than $k\approx 0.3$, because the selection favours extremely star-forming galaxies with high HI content.

We can further disentangle the influence of the galaxy clustering and the intensity maps by considering the scale-dependent cross-correlation coefficient $r(k)$ which can be approximated as
\begin{equation}
r(k)=\frac{\Delta^2_{\rm X}(k)}{\sqrt{\Delta^2_{\rm HI}(k) \Delta^2_{\rm galaxy}(k)}}.
\label{r_def}
\end{equation}
The Poisson noise contribution is removed from the galaxy auto power spectra but not from the intensity mapping power spectrum as reasoned in Sec.~\ref{IMSN}. The Poisson noise contribution to the cross-correlation is caused by optically selected galaxies correlating with the positions of the HI-rich galaxies which is not analytically determinable and therefore incorporated into the correlation coefficient $r(k)$. 

The correlation coefficient of our simulations as a function of $k$ is presented in Fig.~\ref{r_k_model}. The shaded areas mark the coefficients for the three different galaxy selections: blue for star-forming galaxies, red for quiescent galaxies and dark blue for WiggleZ-selected galaxies. In our experimental set-up, $r(k)$ can be larger than 1 since we do not subtract the Poisson noise from the intensity mapping auto power spectrum and the cross-correlations. This noise contribution in the denominator of Equ.~\ref{r_def} is the same for all galaxy populations and models such that it does not bias the comparison. 

The correlation coefficient $r(k)$ for the WiggleZ populations is consistently higher than the other galaxies for all different star formation recipes. We also see that there is a different shape of $r(k)$ for the quiescent and star-forming galaxies, and the star-forming galaxies correlate more strongly with the intensity maps on small scales. The BR model produces the highest cross-correlation coefficients for all galaxy populations, which can be interpreted as producing the tightest correlation between star formation history and HI gas. 

Furthermore in Fig~\ref{r_k_random}, as a null test, we have confirmed our results by cross-correlating the galaxy selections with HI maps where we randomly sampled the HI masses from the given HI mass functions at $z=0.9$. We find that the correlations are close to identical for the different galaxy colours and thereby confirm that the shape of $r(k)$ depends on the relation between star-formation and HI masses for individual objects.

We have quantified the predicted scale-independent galaxy biases $\tilde b$ and correlation coefficient $\tilde r$ for the different models and galaxy populations in Tab.~\ref{table_b}. For scales $k<0.2\hM$, $\tilde r$ is not significantly affected by the selected galaxy type with difference on a level of a few percent. 

The shape of the correlation coefficient is determined by the correlation between the star formation history and to the HI content of galaxies as a function of environment. For example, star-forming galaxies are more weakly clustered on smaller scales than quiescent galaxies, since they tend to reside in less dense regions and outside of clusters where HI content is higher. For cross-correlations, the amplitude of the power spectrum is determined by the overlap of the two datasets, specifically, how star-formation correlates with HI content. We observe a scale-dependent effect on the cross-correlation coefficient which depends on the star-formation history of the galaxy population in association with the HI contents of nearby galaxies. 
\begin{figure}
\includegraphics[width=0.5\textwidth, clip=true, trim= 0 100 0 50]{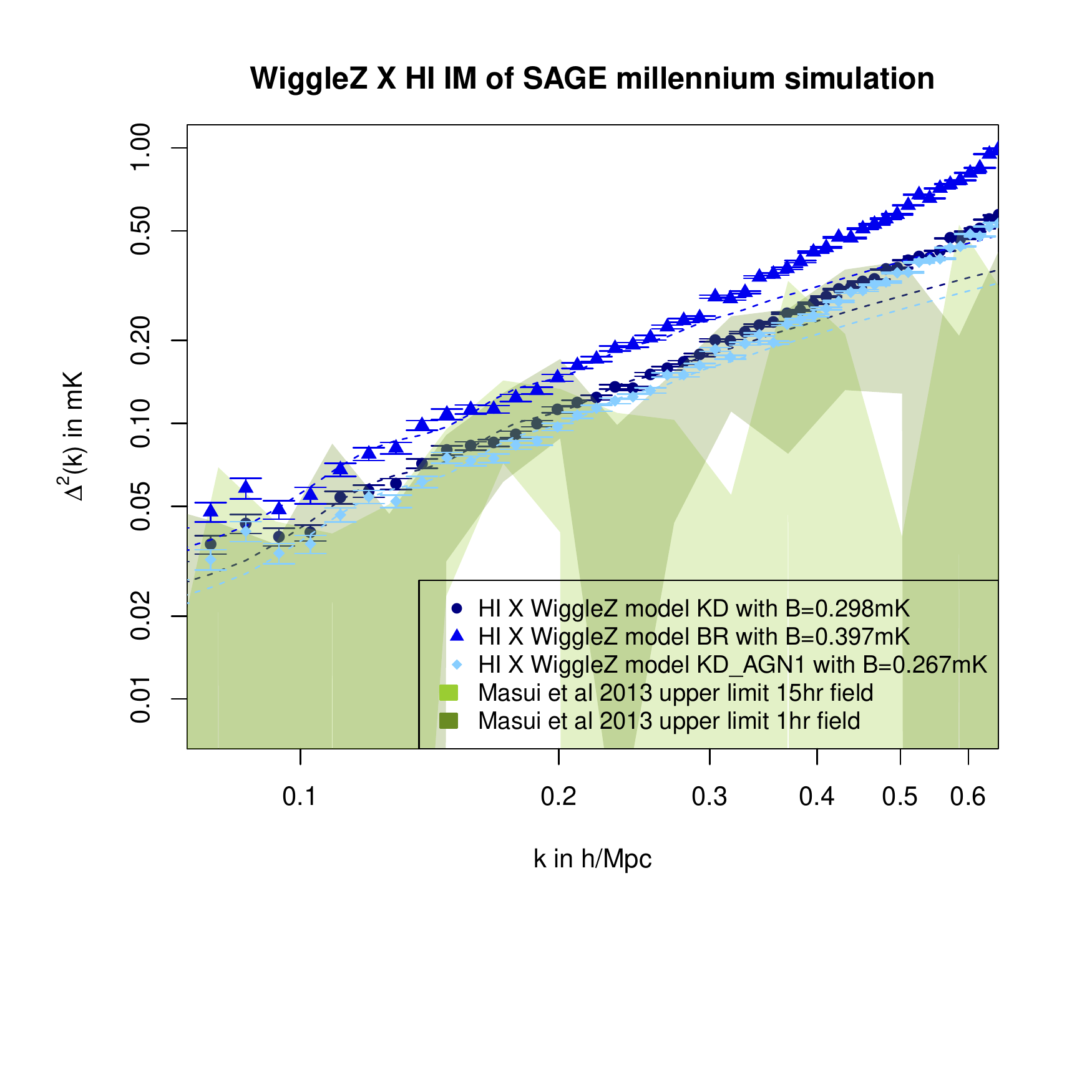}
\caption{Cross power spectrum of WiggleZ selected galaxy populations and intensity maps given in units of mK for all star formation models in blue symbols in comparison to the measurements of the GBT data with WiggleZ galaxies (\citealt{2013ApJ...763L..20M})
 with median redshift of 0.8 indicated by green shaded areas. Note that the factor $B$ given in the legend converts as $B=\bar T_{\rm HI} \tilde b_{\rm HI} \tilde b_{\rm Wig}$}
\label{PS_wigX}
\end{figure}

\begin{itemize}
\item{\textbf{Comparison to GBT results}}\\
In Fig.~\ref{PS_wigX}, we compare the simulated WiggleZ galaxies of the three star-formation models to the measurements of the GBT team published in \cite{2013ApJ...763L..20M} (marked by the green shaded areas). The simulated cross-correlation is close to the upper limit of the errors on the GBT measurements. This is likely due to the same reasons why the WiggleZ simulation over-estimated the observational power spectrum in Fig.~\ref{PS_Wig}, which are that the data is taken at lower redshift  $z=0.8$ rather than 0.905, the power spectrum is estimated in redshift space rather than real space, the Millennium simulation was run with slightly outdated cosmological parameters, i.e. $\sigma_{8}=0.9$, and the GBT intensity maps suffer from significant instrumental systematic effects. Considering these challenges, we believe our model is in reasonable agreement with the data. Preliminary results indicate that the BR model overestimates the power of the cross-correlation and is a less good fit to the HIMF (as seen in Fig.\ref{HIMF}) and data slightly favours the KD SF recipe.
We are planing to extend our model with systematic effects in our intensity mapping cross-correlation simulations in a future study. 
\end{itemize}

\section{Summary and Conclusions}
\label{sec5}
We realistically model the signal of the cross-correlation of intensity mapping observations with optical galaxy surveys at $z\approx 0.9$ using three different star-formation recipes in the SAGE semi-analytical simulations.
We investigate the relation between HI gas and stellar mass as a function of galaxy colour, and find highly star-forming galaxies to lie at in the high-end of the stellar and HI masses, but to exhibit relatively small HI-to-stellar mass ratios. 

The HI density given by our simulation is well within the current observational constraints, and we find that the scale-dependent HI clustering does not depend on the choice of the star formation model. The galaxy power spectrum confirms that quiescent (red) galaxies exhibit more clustering power on smaller scales than do star-forming galaxies. Moreover the cross-correlation power spectrum of the intensity maps with optically selected galaxy populations demonstrates how the star-formation history of the galaxies can influence the clustering on smaller scales. 
The shape of the cross power spectrum of quiescent and star-forming galaxies is indistinguishable. However, when considering the cross-correlation coefficient, a tighter correlation between the blue population and intensity maps is seen.

The highly star-forming galaxies selected by a WiggleZ like colour cut show a much higher cross-correlation power with the intensity maps on small scales ($k>0.3\hM$) than red or blue galaxies. This confirms that the high HI content of the WiggleZ selected galaxies correlates more strongly with the intensity maps dominated by HI rich galaxies, and that this effect is clearly distinguishable in the power spectrum. 

We calculate the scale-independent coefficient $\tilde r$ on large scales and show that $\tilde r$ has only a marginal model dependence for the different star-formation recipes, with a maximum range of $10\%$. The scale-dependence of the cross-correlation coefficient $r(k)$ exhibits a higher amplitude of for WiggleZ like  galaxies for all scales, motivating the cross-correlation of intensity maps with UV-selected, highly star-forming galaxies.

The conclusions from this study are:
\begin{itemize}
\item We have set up a simulation to model the cross-correlation of optical galaxy surveys for quiescent and star-forming populations with present and future intensity mapping experiments. The model can be used for the data analysis of the GBT telescope and future Parkes observations. The predictions for \ObHI~are within the observational constraints. The comparison with data slightly favours the Krumholz-Dekel star-formation recipe. 
\item On large scales $k\sim 0.2\hM$, the power of the cross-correlation changes only marginally (less than 10\%) with star-formation recipe and galaxy cut. For near-future experiments these differences lie within the measurement uncertainties.
\item We find that the shape of cross-correlation coefficient needs to be considered in data analysis on scales $k>0.3\hM$. Additionally, the cross-correlation coefficient shape depends on the star formation history and the HI content of the optically selected galaxies as a function of environment. This could be used, for example, in an experiment to measure the relative HI content of two independent, optically selected galaxy populations by the comparison of their cross-correlations with intensity maps. This could be done for high redshifts where measurements of gas properties are not feasible with present instruments. 
\end{itemize}
\section*{Acknowledgments}
We would like to thank Darren Croton for the use of the semi-analytical model and his useful comments throughout the project. CB acknowledges the support of the Australian Research Council through the award of a Future Fellowship.
The Centre for All-Sky Astrophysics is an Australian Research Council Centre of Excellence, funded by grant CE11E0090.

\bibliographystyle{mn2e}
\bibliography{library}

\label{lastpage}

\end{document}